\documentclass[aps, longbibliography, twocolumn]{revtex4-1}
\usepackage{bm}
\usepackage{graphicx}
\usepackage{amsmath}
\usepackage{amssymb} 
\usepackage[utf8]{inputenc}
\usepackage[T1]{fontenc}
\usepackage{color}
\usepackage{xcolor} 
\usepackage{upgreek}
\usepackage{subfigure}
\usepackage[unicode=true,colorlinks=true,citecolor=blue]{hyperref} 
\usepackage{placeins}
\usepackage{lipsum}
\usepackage{epsfig}
\usepackage{wrapfig} 
\usepackage[normalem]{ulem}
\usepackage{units}
\usepackage{cancel}
\newcommand{\nix}[1]{}

\newcommand{\pderiv}[2]{\frac{\partial #1}{\partial #2}}
\renewcommand{\phi}{\varphi}

\renewcommand{\i}{\mathrm i}
\newcommand{\e}{\mathrm e}

\newcommand{\beq}{\begin{equation}}
\newcommand{\eeq}{\end{equation}}
\newcommand\beqa{\begin{eqnarray}}
\newcommand\eeqa{\end{eqnarray}}
\newcommand\ba{\begin{array}}
\newcommand\ea{\end{array}}

 \newcommand{\aver}[1]{\left \langle #1 \right \rangle}

\newcommand{\psum}{\sum\limits_{\boldsymbol{p}}}

\begin{document}

\title{Photocurrents induced by structured light}

\author{A.\,A.\,Gunyaga, M.\,V.\,Durnev, S.\,A.\,Tarasenko}

\affiliation{Ioffe Institute, 194021 St. Petersburg, Russia}

\begin{abstract}
Advances in manipulating the structure of optical beams enable the study of interaction between structured light and low-dimensional semiconductor systems. We explore the photocurrents in two-dimensional systems excited by such inhomogeneous radiation with structured electromagnetic field. Besides the contribution associated with the intensity gradient, the photocurrent contains contributions driven by the gradients of the Stokes polarization parameters and the phase of the electromagnetic field. 
We develop a microscopic theory of the non-linear non-local intraband transport of electrons induced by  
electromagnetic field of structured radiation and derive analytical expressions for the photocurrent contributions. 
The theory is applied to analyze the radial and azimuthal photocurrents excited by twisted radiation beams carrying orbital angular momentum, and possible experiments to detect the photocurrents are discussed.
\end{abstract}
  
\maketitle

\section{Introduction}
\label{introduction}

Structured light, extending from intensity or polarization gratings to beams carrying orbital angular momentum 
and fields with fully controlled spatio-temporal structure, has a variety of application in spectroscopy, metrology, quantum information processing, etc.~\cite{Forbes2021, Allen1999, MolinaTerriza2007, Knyazev2018}. 
While significant progress has been achieved recently in the optical methods of structuring the light beams~\cite{Maurer2007, Wei2015, Choporova2017,  Zhifeng2020} and controlling the optical forces acting on micro- and nano-scale dielectric particles~\cite{Ashkin2000, Andrews_book2011, Marago2013, Urban2014, Mantsevich2017, Yang2021}, less is known so far about the interaction of structured light beams with charge carriers in 
low-dimensional semiconductor systems. The latter research area, bridging advances in optics and solid state physics~\cite{Ji2020, Sederberg2020, Lai2022,Bhattacharya2022}, is intriguing from fundamental point of view and crucial for the development of optoelectronics.

The electron-photon interaction in semiconductors is sensitive to the local field polarization and the local symmetry breaking~\cite{OO_book,Hubmann2020}. 
This underlies the physics of photogalvanic effects, where the ac electric field of linearly or circularly polarized radiation 
drives dc electric currents in non-centrosymmetric structures~\cite{Sturman_book, Ivchenko_book, Sipe2000, Tarasenko2011, Durnev2019, Candussio2020, Durnev2021b,Steiner2022, Leppenen2023, Parafilo2022}. The key feature of structured light is that the electromagnetic field parameters, such as polarization or phase, vary at the scale of light wavelength. This length is typically much larger than 
the scale of non-locality in semiconductors, which makes it challenging to study the photoresponse associated with the light structure.
Nevertheless, several groups have recently reported the observation of the photocurrents induced by the Laguerre-Gaussian beams 
and sensitive to the orbital angular momentum of light~\cite{Ji2020, Sederberg2020, Lai2022}. These advances open a new page in the research of photoelectric phenomena in solids and stimulate theoretical studies of the photocurrents 
induced by structured light. Theoretical results in the field are limited so far to the phenomenological analysis of the photocurrents~\cite{Ji2020,Lai2022}, numerical quantum-mechanical calculations of non-stationary currents  
in quantum rings and dots~\cite{Quinteiro2009b,Waetzel2016,Waetzel2020}, and the quantum-mechanical calculation of the rotational photon drag~\cite{Quinteiro2010}. The latter belongs to the family of photon drag effects being extensively studied in different material systems and geometries~\cite{Danishevskii1970, Gibson1970, Perel1973, Luryi1987,Shalygin2007,Hatano2009,Karch2010,Entin2010,Stachel2014,Obraztsov2014, Glazov2014,Plank2016,Mikheev2018,Shi2021}.

Here, we develop a microscopic theory of photocurrents induced in two-dimensional electron systems by structured radiation. We explore what types of direct currents
can be driven by spatially inhomogeneous electromagnetic waves acting upon free carriers.
It is found that, besides the contribution associated with the intensity gradient, the photocurrent contains contributions driven by the gradients of the Stokes polarization parameters and the phase of the electromagnetic field. In particular, the photocurrent emerging between the domains excited by left-handed and right-handed circularly polarized radiation can be interpreted as the chiral edge current between the photo-induced topological phases with the opposite Floquet-Chern numbers~\cite{Kitagawa2011,Lindner2011}.
In the framework of Boltzmann kinetic theory, we derive analytical expressions for all the photocurrent contributions corresponding to the intraband transport of electrons.

\begin{figure}[b]
\includegraphics[width=0.98\linewidth]{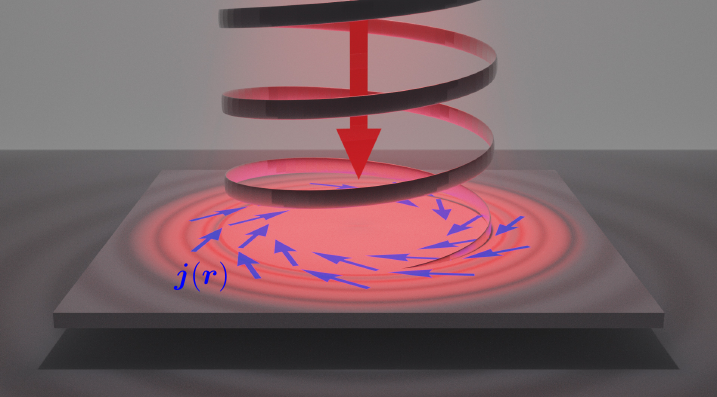}
\caption{Photocurrents induced by structured light. Inhomogeneous ac electric field of twisted radiation beam generates dc electric current $\boldsymbol{j}(r)$ in two-dimensional electron gas.}
\label{fig1}
\end{figure}

The developed theory is applied then to analyze the spatial distribution of the photocurrents induced by twisted-light beams, see Fig.~\ref{fig1}. In this geometry, the emergent photocurrent $\bm j (\bm r)$ has both radial and azimuthal components which are controlled by the parameters of the beam. The radial photocurrents lead to a redistribution of electric charge in the two-dimensional plane and form the radial photovoltage. The azimuthal photocurrents induce a static magnetic field and the corresponding magnetization. We calculate the photovoltage and the static magnetic field and discuss their dependence on the beam angular momentum and polarization.

\section{Kinetic theory}

We consider a two-dimensional electron gas (2DEG) in the $xy$ plane (at $z = 0$) which is irradiated by a spatially-inhomogeneous monochromatic electromagnetic wave. In the 2DEG plane, the electric field has the form 
\begin{equation}\label{E_field}
\bm E (\bm r, t) = \bm E (\bm r) \mathrm{e}^{-\mathrm{i}\omega t}+\mathrm{c.c.} \,,
\end{equation}
where $ \bm E (\bm r)$ is the field amplitude and $\bm r = (x, y)$ is the in-plane coordinate. Generally, because of screening, the spatial profile of
the electric field acting upon the electrons differs from that of the incident wave. The total field can be calculated in the framework of the Lindhard screening theory~\cite{Lindhard1954,GiulianiVignale_book2005}. However, for the amplitude $\bm E (\bm r)$ which varies smoothly in the 2DEG plane, i.e., in the long wavelength limit, screening is negligible.
This regime reliably holds for electromagnetic fields of radio-frequency and terahertz spectral ranges with the inhomogeneity scale 
of the order of the wavelength~\footnote{See paragraph after Eq.~\eqref{currentnew} for details}.  
Therefore, we neglect the difference between the profiles of the incident and actual electric fields. 

The electromagnetic field necessarily contains the magnetic component $\bm B(\bm r, t) = \bm B (\bm r) \mathrm{e}^{-\mathrm{i}\omega t}+\mathrm{c.c.}$ with the out-of-plane projection
\begin{equation}\label{Bz_field}
B_z = -\i  \frac{c}{\omega} \left( \pderiv{E_y}{x} - \pderiv{E_x}{y} \right) \,,
\end{equation}
which follows from the Maxwell equation. In the model of 2D transport, the electrons interact only with the in-plane component $\bm E_{\parallel} = (E_x, E_y)$ of the electric field and the out-of-plane component $B_z$ of the magnetic field.  

The electron kinetics in the presence of electric and magnetic fields is described by the Boltzmann equation
\begin{equation}\label{boltzman}
    \pderiv{f}{t} + \bm v \cdot \nabla f +e \left[ \bm{ E}_{\parallel} (\bm r, t) + \frac{1}{c} \bm v \times \bm B (\bm r, t) \right] \cdot \pderiv{f}{\bm p}= I \{ f \} \,,
\end{equation}
where 
$f(\bm p, \bm r, t)$ is the electron distribution function, $\bm p = (p_x, p_y)$ and $\bm v = \bm p/m^*$ are the momentum and velocity, respectively, $\nabla = \partial / \partial \bm r$ is the 2D nabla operator, $e$ is electron charge, $m^*$ is the effective mass, and $I \{ f \}$ is the collision integral. In what follows, we do calculations in the relaxation time approximation and take the collision integral in the form
\begin{equation}\label{collision_integral}
 I \{ f \} = - \frac{f - \aver{f}}{\tau}  \,,
\end{equation}
where $\aver{f}$ is the distribution function averaged over the directions of $\bm p$, and $\tau$ is the relaxation time.

Dc electric currents emerge in the second order in the ac field amplitude. Therefore, we solve Eq.~\eqref{boltzman} iteratively by 
expanding the distribution function $ f(\bm p, \bm r, t)$ in the series in the field amplitude as follows
\begin{equation}\label{Eseries}
    f(\bm p, \bm r, t) = f_0+\left[f_1 (\bm p, \bm r) \mathrm{e}^{-\mathrm{i}\omega t}+\mathrm{c.c.}\right]+f_2 (\bm p, \bm r),
\end{equation} 
where $f_0(\varepsilon)$ is the equilibrium distribution function, $\varepsilon = p^2/2m^*$ is the electron energy, $f_1 \propto E$ is the first-order correction, and $f_2$ is the stationary second-order correction which determines the dc current. The correction $f_2$ contains contributions $\propto E E^*$ and $\propto E B^*, B E^*$. Other second-order corrections, such as $\propto E E$ and $\propto E B$, oscillate at $2\omega$ and do not contribute to the dc current.

The functions $f_1$ and $f_2$ satisfy the differential equations, which follow from Eq.~\eqref{boltzman},
\begin{align}
    -\mathrm{i}\omega f_1 + \bm v \cdot \nabla f_1 + e \bm{E}_{\parallel} (\bm r) \cdot \frac{\partial f_0}{\partial\bm p} = I \{ f_1 \} \:, \label{f1} \\
   \bm v \cdot \nabla f_2+\left\{ e \left[ \bm{E}_{\parallel} (\bm r) + \frac{1}{c} \bm v \times \bm B (\bm r) \right] \cdot \frac{\partial f_1^*}{\partial\bm p}+\mathrm{c.c.}\right\} = I \{ f_2 \} \label{f2}\:.
\end{align}
The term with the Lorentz force in Eq.~\eqref{f1} vanishes because $\partial f_0/ \partial \bm p \propto \bm p$ and $\bm v \times \bm B$ is orthogonal to $\bm p$.

The local density of the dc electric current is given by
\begin{equation}
    \boldsymbol{j} (\bm r) = e \nu \sum\limits_{\boldsymbol{p}}\boldsymbol{v} f_2(\bm p, \bm r) \,, \label{current}
\end{equation}
where $\nu$ is the factor of spin/valley degeneracy. Multiplying Eq.~\eqref{f2} by $\bm v$, summing up the result over $\bm p$, and 
integrating the term $\propto \partial f_1^* / \partial \bm p$ by parts one obtains
\begin{multline}
    \bm j (\bm r)  = - e \nu \tau \sum_{\bm p} \bm v (\bm v \cdot \nabla f_2)
    + \frac{e^2 \nu \tau }{m^*} \left[ \bm{E}_{\parallel} (\bm r) \sum_{\bm p} f_1^* + \mathrm{c.c.} \right] \\
     - \frac{e^2 \nu \tau}{m^* c} \left[ \bm B(\bm r) \times \sum_{\bm p} \bm v f_1^* + \mathrm{c.c.} \right]\label{currentnew} \, .
\end{multline}

To proceed further we take into account that the field is smoothly varying in the 2DEG plane, i.e., the scale of the field variation $L$ 
is much larger than the mean free path of electrons $l = v_F \tau$ and $v_F/\omega$, where $v_F$ is the Fermi velocity. 
Assuming that $L \gtrsim \lambda$, where $\lambda = 2\pi c/\omega$ is the wavelength of the incident field, the above conditions  
reduce to $l \ll \lambda$ and $v_F \ll c$. The former inequality holds very well for radio-frequency and terahertz fields even in 
high-mobility systems (1 THz corresponds to $\lambda \sim 0.3~$mm). The latter inequality is automatically fulfilled for conduction-band electrons in solids. 
We also note that the electric field screening by 2DEG is negligible if $L \gg  (2\pi \sigma/c) \lambda$, where $\sigma$ is the 2D conductivity.
This condition can be readily fulfilled, particularly in the paraxial approximation of the incident radiation. However, the effects of screening can play an important role if $L \ll \lambda$, which can be achieved, e.g., in plasmonic structures where the electric field is enhanced near small metal particles 
located at the 2DEG plane.

For smoothly varying fields, the gradients of the field components and the distribution function are small and can be treated perturbatively. 
We do such calculations for the photocurrent Eq.~\eqref{currentnew} in the first order of $l/L$. 

The third term in Eq.~\eqref{currentnew} contains 
$B_z(\bm r)$, which is already determined by the field gradients, see Eq.~\eqref{Bz_field}, 
therefore, the sum $\sum_{\bm p} \bm v f_1^*$ can be calculated in the local response approximation.
Multiplying Eq.~\eqref{f1} by $\bm v$, summing up the resulting equation over $\bm p$, and neglecting 
the gradient term $\propto \nabla f_1$ one obtains
\begin{equation}\label{SFA}
 \sum_{\bm p}\bm v f_1 = \frac{\sigma}{e \nu} \bm E_{\parallel} (\bm r) \,,  
\end{equation}
where $\sigma$ is the conductivity, 
\begin{equation}
   \sigma = \frac{n e^2\tau}{m^* (1-\mathrm{i}\omega\tau)} \,,  
\end{equation}
and $n = \nu \sum_{\bm p}f_0$ is the 2D electron density. 

The sum $\sum_{\bm p} f_1^*$ in the second term in Eq.~\eqref{currentnew} 
can be calculated by summing up Eq.~\eqref{f1} over $\bm p$ and using Eq.~\eqref{SFA}, which gives
\begin{equation} \label{charge}
    \psum f_1=-\frac{\mathrm{i}\sigma}{e \nu \omega} \nabla \cdot \bm{E}_{\parallel}(\bm r) \,,
\end{equation}
where $\nabla \cdot \bm{E}_{\parallel} = \partial E_x / \partial x + \partial E_y / \partial y$. 

The first term in Eq.~\eqref{currentnew} is determined by the gradient of $f_2$, 
therefore the sums $\sum_{\bm p} v_\alpha v_\beta f_2$ can be also calculated in the local response approximation. 
The function $f_2$ contains the zero and second angular harmonics which both contribute to the sums $\sum_{\bm p} v_\alpha v_\beta f_2$ 
since 
\begin{equation}
\sum_{\bm p} v_\alpha v_\beta f_2 = \sum_{\bm p} \frac{v^2}{2} \delta_{\alpha\beta} f_2 \,  + \sum_{\bm p} \left(v_\alpha v_\beta - \frac{v^2}{2} \delta_{\alpha\beta} \right) f_2 \,.
\end{equation}
The contribution of the zero angular harmonic is calculated as follows 
\begin{equation}
\label{K}
\sum_{\bm p}  \frac{v^2}{2} f_2 = \frac{K}{m^* \nu}  \,,
\end{equation}
where $K = \nu \sum_{\bm p} (m^* v^2 /2) f_2$ is the density of the excess kinetic energy of electrons.
If the energy relaxation of electrons is faster than the energy diffusion, the excess energy $K$ can be found 
from the balance between the energy gain from the high-frequency field $2 (\mathrm{Re} \, \sigma) |\bm E_{\parallel}|^2$ and the energy dissipation  $K/\tau_\varepsilon$, where $\tau_\varepsilon$ is the energy relaxation time~\footnote{To describe the energy relaxation of electrons one needs to go beyond the collision integral~\eqref{collision_integral}}. Thus, the energy balance gives  
$K = 2 \tau_\varepsilon (\mathrm{Re} \, \sigma) |\bm E_{\parallel}|^2$.
To calculate the contributions of the second angular harmonics, we multiply Eq.~\eqref{f2} 
by $v_x v_y$ or $v_x^2 - v_y^2$ and sum up the result over $\bm p$, which yields 
\begin{eqnarray}\label{2nd}
    \psum v_{x}v_{y} f_2 &=& \frac{2 \tau \mathrm{Re} \, \sigma}{m^* \nu} \left( E_{x} E_{y}^* + E_{x}^* E_{y} \right)\:, \\
    \psum \frac{v_x^2 - v_y^2}{2} f_2 &=& \frac{2 \tau \mathrm{Re} \, \sigma}{m^* \nu} \left( |E_{x}|^2 - |E_{y}|^2 \right) \nonumber\:.
\end{eqnarray}

Finally, combining all the contributions to the photocurrent, we obtain
\begin{equation}\label{j_total}
\bm j =  \bm j^{\rm (th)}  + \bm j^{\rm (pol)}  + \bm j^{\rm (ph)}  \,,
\end{equation}
where 
\begin{equation}\label{jxyth}
\bm j^{\rm (th)} =  - 2\frac{e \tau\tau_\varepsilon \, \mathrm{Re} \, \sigma}{m^*} \nabla S_0 \,,
\end{equation}
\begin{eqnarray}\label{jxypol}
j_x^{\rm (pol)} &=& - \frac{e \tau^2 \, \mathrm{Re} \, \sigma}{m^*} \left( \pderiv{S_1}{x} + \pderiv{S_2}{y} - \frac{1}{\omega\tau} \pderiv{S_3}{y} \right) , \nonumber \\ 
j_y^{\rm (pol)} &=& - \frac{e \tau^2 \, \mathrm{Re} \, \sigma}{m^*} \left(\pderiv{S_2}{x} -\pderiv{S_1}{y} + \frac{1}{\omega\tau} \pderiv{S_3}{x} \right) ,
\end{eqnarray}
\begin{equation}\label{jxyph}
\bm j^{\rm (ph)} = - 2\frac{e \tau \, \mathrm{Re} \, \sigma}{m^* \omega} \mathrm{Im} \left( E_x \nabla E_x^* + E_y \nabla E_y^* \right) ,
\end{equation}
$\mathrm{Re} \, \sigma = n e^2 \tau /[m^* (1+\omega^2 \tau^2)]$, $S_0 = |\bm E_{\parallel}|^2$, 
$S_1 = |E_{x}|^2-|E_{y}|^2$, $S_2 = E_{x} E_{y}^*+ E_{x}^* E_{y}$, and $S_3 = \i \left( E_{x} E_{y}^*- E_{x}^* E_{y} \right)$.
In the paraxial approximation for normally incident radiation, $S_0$, $S_1$, $S_2$, and $S_3$ correspond to the (non-normalized) 
Stokes parameters. The parameter $S_0$ describes the radiation intensity, $S_1 / S_0$ and $S_2 / S_0$ are the degrees of linear polarization in the $(xy)$ axes and in the diagonal axes, respectively, and $S_3 / S_0$ is the degree of circular polarization. All the above contributions to the dc current are proportional to the square of the electric field amplitude, i.e., the radiation intensity, and, therefore, belong to the class of photocurrents.

Besides the photothermoelectric current $\bm j^{\rm (th)}$ originating from inhomogeneous heating and proportional to the gradient of the radiation intensity, the photocurrent contains 
the contributions $\bm j^{\rm (pol)}$ and $\bm j^{\rm (ph)}$. 
The first one, $\bm j^{\rm (pol)}$ given by Eqs.~\eqref{jxypol}, is determined by the gradients of the (non-normalized) polarization Stokes parameters $S_j$ ($j = 1,2,3$).
This photocurrent is induced by the electromagnetic field with spatially varying polarization, see Fig.~\ref{Fig_polarization}, and emerges 
even if the field intensity is constant across the 2D gas plane. It is also induced by the field with a fixed polarization but non-uniform intensity, e.g., when the ratio $S_j/S_0$ is constant but $S_0$ depends on $\bm r$. In this particular case, the photocurrent $\bm j^{\rm (pol)}$ can be seen as proportional to the in-plane gradient of the radiation intensity, similarly to the photothermoelectric contribution. However, its direction is determined by the field polarization and it can flow both along or perpendicularly to the intensity gradient. The polarization-sensitive photocurrent flowing perpendicularly to the intensity gradient is somewhat similar to the edge photogalvanic current in 2DEG~\cite{Candussio2020,Durnev2023}.

The contribution $\bm j^{\rm (ph)}$ given by Eq.~\eqref{jxyph} does not even require polarization gradients to emerge. 
For the field $\bm{E}(\bm r) = \bm E_0 \e^{i \phi(\bm r)}$ with the constant amplitude and polarization,   
the photocurrent $\bm j^{\rm (ph)}$ is proportional to the gradient of the phase $\phi(\bm r)$.
An example of such a field $\bm{E}(\bm r)$ is the plane electromagnetic wave obliquely incident on the 2D system.
In this geometry, the phase varies linearly in 2D plane as $\phi = \bm q_{\parallel} \cdot \bm r$, where $\bm q_{\parallel}$ is the in-plane component of the wavevector. 
The emerging photocurrent $\bm j^{\rm (ph)} \propto \bm q_{\parallel} |\bm E_{\parallel}|^2$ is proportional to the in-plane wavevector  
$\bm q_{\parallel} $ and corresponds to the photon drag effect~\cite{Karch2010,Stachel2014,Glazov2014}. Thus, the phase sensitive contribution $\bm j^{\rm (ph)}$ given by Eq.~\eqref{jxyph} can be viewed as a generalization of the photon drag effect to the electromagnetic field with arbitrary varying phase.     

All the photocurrent contributions discussed above are proportional to the cube of the electric charge $e$. It means that the photocurrents excited in
$n$-type and $p$-type structures are directed oppositely while the flows of carriers, electrons or holes, are co-directed.

Equation~\eqref{j_total} with the contributions~\eqref{jxypol} and \eqref{jxyph} is the main result of our work.  It describes the generation of dc current in 2DEG by arbitrary spatially inhomogeneous electromagnetic field. Below, we apply it to study the photocurrents induced by fields with polarization gradients and 
by beams of twisted light.

\section{Photocurrents induced by gradients of field polarization}

\begin{figure*}
\includegraphics[width=0.95\linewidth]{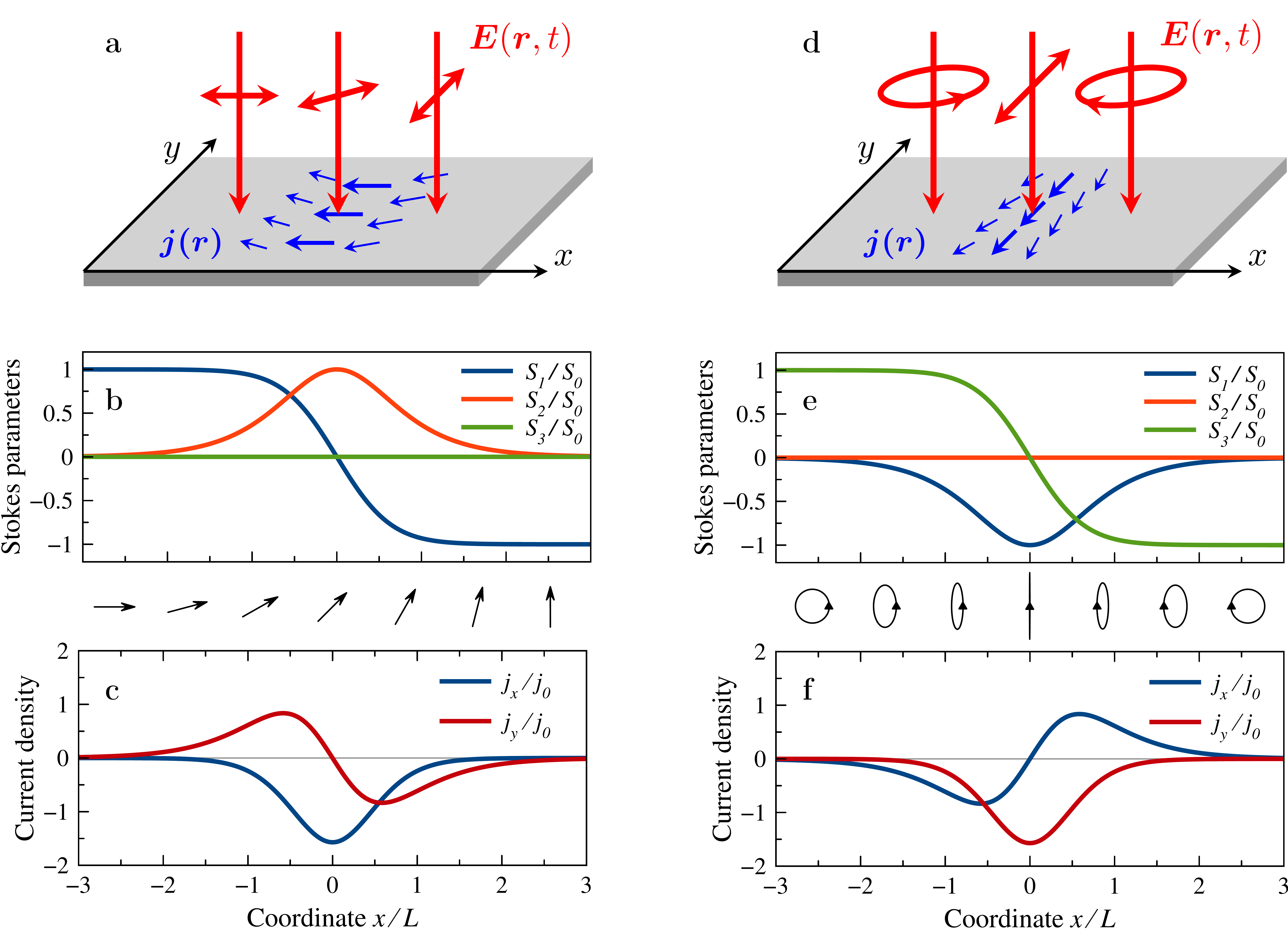} 
\caption{AC electromagnetic field $\bm {E}(\bm r ,t)$ with
spatially inhomogeneous polarization induces direct photocurrents 
$\boldsymbol{j}(\boldsymbol{r})$ in the regions where the polarization varies. (a)-(c) Photocurrents induced by linearly polarized radiation whose polarization vector turns in the electron gas plane. (b) Spatial profiles of the Stokes parameters $S_j(x)/S_0$ of the incident radiation. The in-plane field polarizations are shown by black arrows. (c) Spatial distribution of the $x$ and $y$ components of the photocurrent density $\bm j (x)$. 
(d)-(f) Photocurrents induced by radiation with the polarization  varying from left-handed to right-handed circular polarization. (e) Spatial profiles of the Stokes parameters $S_j(x)/S_0$ of the incident radiation. The field polarizations are shown by black arrows and ellipses. (f) Spatial distribution of the $x$ and $y$ components of the photocurrent density $\bm j (x)$ at $\omega \tau = 1$.}
\label{Fig_polarization}
\end{figure*}

Consider the photocurrents induced by electromagnetic fields with the constant in-plane amplitude $|\bm E_{\parallel}|$ in the 2DEG plane 
and the polarization varying along the $x$ axis. Figures~\ref{Fig_polarization}(a) and~\ref{Fig_polarization}(d) show examples of such fields. In Fig.~\ref{Fig_polarization}(a), the field $\bm E_{\parallel}$ is linearly polarized and the polarization rotates from $\bm E_{\parallel} \parallel x$ at large negative $x$ to  
$\bm E_{\parallel} \parallel y$ at large positive $x$. In Fig.~\ref{Fig_polarization}(d), the field polarization varies from the left-handed to the right-handed 
circular polarization through the linear polarization.
  
Figures~\ref{Fig_polarization}(b) and~\ref{Fig_polarization}(c) show the spatial dependence of the Stokes parameters and the spatial distribution of the emergent photocurrent density 
$\bm j(x)$ for the radiation sketched in Fig.~\ref{Fig_polarization}(a). The polarization parameter $S_1 / S_0$ varies from $+1$ at $x \rightarrow -\infty$ to $-1$  at $x \rightarrow +\infty$ whereas $S_2/S_0$ vanishes at $x \rightarrow \pm \infty$ and reaches its maximum value of 1 at $x = 0$. 
We take the spatial profile of the electric field amplitude in the form 
$E_x = E_0 \cos \Phi(x)$, $E_y = E_0 \sin \Phi(x)$,
where $\Phi(x) = (\pi / 4) [\mathrm{tanh}(x/L) + 1]$, which corresponds 
to the Stokes parameters $S_0 = E_0^2$, $S_1 = E_0^2 \cos 2\Phi(x)$, $S_2 = E_0^2 \sin 2\Phi(x)$, and $S_3 = 0$.
As follows from Eqs.~\eqref{jxyth}-\eqref{jxyph}, the currents $\bm j^{\rm (th)}$ and $\bm j^{\rm (ph)}$ vanish for such a field, and
the photoresponse is determined by the polarization sensitive contribution $\bm j^{\rm (pol)}$. The current $\bm j^{\rm (pol)}$ emerges
in the region where $S_1$ and $S_2$ vary, with $j_x ^{\rm (pol)} \propto \partial S_1 / \partial x$ and $j_y ^{\rm (pol)} \propto \partial S_2 / \partial x$. As a result, the photocurrent $\bm j $ has both components
\begin{eqnarray}
&& j_x  (x) = - \frac{\pi}{2} \frac{\cos [(\pi /2) \mathrm{tanh}(x/L)]}{\cosh^2 (x/L)} j_0 \,, \nonumber \\
&& j_y  (x) = - \frac{\pi}{2} \frac{\sin [(\pi /2) \mathrm{tanh}(x/L)]}{\cosh^2 (x/L)} j_0 \,,
\end{eqnarray}
where $j_0 = - n e^3 \tau^3 E_0^2 / [m^{*2} L (1+\omega^2\tau^2)]$. The distributions $j_x (x)$ and $j_y (x)$ are plotted in Fig.~\ref{Fig_polarization}(c).
On average, the carriers (electrons with $e<0$ or holes with $e>0$) 
flow from the domain with $\bm E_{\parallel} \parallel x$ to the domain with $\bm E_{\parallel} \parallel y$.

Figures~\ref{Fig_polarization}(e) and~\ref{Fig_polarization}(f) show the spatial distributions of the Stokes parameters and the photocurrent density  
for the radiation with varying helicity as sketched in Fig.~\ref{Fig_polarization}(d). Here, the profile of the electric field amplitude 
is taken in the form $E_x = \i E_0 \sin \Phi(x)$, $E_y = E_0 \cos \Phi(x)$,
where $\Phi(x) = (\pi / 4) \mathrm{tanh}(x/L)$. The corresponding Stokes parameters have the form
$S_0 = E_0^2$, $S_1 = - E_0^2 \cos 2\Phi(x)$, $S_2 = 0$, and $S_3 = - E_0^2 \sin 2\Phi(x)$. 
For this electromagnetic field, the photocurrent $\bm j$ is also solely determined by the polarization 
sensitive contribution $\bm j^{\rm (pol)}$ with the projections $j_x^{\rm (pol)} \propto \partial S_1 / \partial x $ 
and $j_y^{\rm (pol)}  \propto \partial S_3 / \partial x$. Thus, the spatial distributions of the photocurrent components 
are given by the functions
\begin{align}
& j_x  (x) = \frac{\pi}{2} \frac{\sin [(\pi /2) \mathrm{tanh}(x/L)]}{\cosh^2 (x/L)} \, j_0 \,, \nonumber \\
& j_y  (x) = - \frac{\pi}{2} \frac{\cos [(\pi /2) \mathrm{tanh}(x/L)]}{\cosh^2 (x/L)} \frac{j_0}{\omega \tau} \,,
\end{align}  
which are plotted in Fig.~\ref{Fig_polarization}(f).  

Interestingly, the photocurrent between the domains 
with left-handed and right-handed circular polarizations flows along the boundary of the domains, 
as sketched in Fig.~\ref{Fig_polarization}(d). The total boundary current 
\begin{equation}
\label{Jy}
J_y = \int j_y(x) dx = - \frac{n e^3 \tau^2 [S_3(+\infty) - S_3(-\infty)]}{m^{*2} \omega (1+\omega^2\tau^2)}
\end{equation}
does not depend on the domain boundary structure nor the boundary width $L$ and 
is determined by the difference of the Stokes parameter $S_3$ in the domains.
Moreover, in the collisionless limit, i.e., at $\omega \tau \gg 1$, the current $J_y$ is independent of the relaxation time $\tau$.
These features allow us to attribute the photocurrent $J_y$ to the chiral edge current emerging between  
the photo-induced topological phases with the opposite Floquet-Chern numbers~\cite{Kitagawa2011,Lindner2011}.

The photocurrent~\eqref{Jy} is estimated as $J_y \approx 20~\mu$A for the carrier density $n = 5\times10^{11}$~cm$^{-2}$, the relaxation time $\tau = 1$~ps, the effective mass $m^* = 0.03~m_0$, where $m_0$ is the free-electron mass, which correspond to bilayer graphene~\cite{Candussio2020}, $\omega \tau = 1$, and the electric field amplitude $E_\parallel = 0.25$~kV/cm corresponding to the terahertz radiation intensity $I = 1$~kW/cm$^2$.

\section{Photocurrents induced by twisted radiation}

Now, we apply the developed theory to calculate the photoresponse of 2DEG to the twisted radiation, i.e., the electromagnetic waves carrying orbital angular momentum, Fig.~\ref{fig1}. 
Such calculations can be conveniently done in the polar coordinate frame with the radial $\bm e_r = (\cos \phi, \sin \phi)$ 
and azimuthal $\bm e_\phi = (-\sin \phi, \cos \phi)$ unit vectors, where $\phi$ is the polar angle counted from the $x$ axis.
In this frame, the photocurrents $\bm j^{\rm (th)}$,  $\bm j^{\rm (pol)}$, and $\bm j^{\rm (ph)}$ are given by
\begin{equation}\label{jpolarth}
\bm j^{\rm (th)} =  - 2\frac{e \tau\tau_\varepsilon \, \mathrm{Re} \, \sigma}{m^*} \nabla P_0 \,,
\end{equation}
\begin{eqnarray}\label{jpolarpol}
j_r^{\rm (pol)} = -\frac{e \tau^2 \, \mathrm{Re} \, \sigma}{m^{*}} 
\left( \pderiv{P_1}{r}  +\frac{2P_1}{r} + \frac{1}{r}\pderiv{P_2}{\varphi} - \frac{1}{\omega\tau}\frac{1}{r}\pderiv{P_3}{\varphi} \right) , \nonumber \\
j_{\varphi}^{\rm (pol)} = \frac{e \tau^2 \, \mathrm{Re} \, \sigma}{m^{*}} 
\left( \frac{1}{r}\pderiv{P_1}{\varphi} - \pderiv{P_2}{r} - \frac{2 P_2}{r} - \frac{1}{\omega\tau} \pderiv{P_3}{r}  \right) ,  \;\;\;\;\;\;\;
\end{eqnarray} 
\begin{equation}\label{jpolarph}
\bm j^{\rm (ph)} = - 2\frac{e \tau \, \mathrm{Re} \, \sigma}{m^{*}\omega} 
\left[ \mathrm{Im}\left( E_r \nabla E_r^* + E_{\varphi} \nabla E_{\varphi}^*\right) 
+ P_3 \frac{\bm e_\phi}{r}  \right] , \
\end{equation} 
where $\nabla = \bm e_r \partial_r + (\bm e_\phi/r) \partial_\phi$, $P_0 = |E_r|^2 + |E_{\varphi}|^2 = S_0$, 
$P_1 = |E_{r}|^2-|E_{\varphi}|^2$, $P_2 = E_{r} E_{\varphi}^*+ E_{r}^* E_{\varphi}$,
 and $P_3 = \i \left(E_{r} E_{\varphi}^*- E_{r}^* E_{\varphi} \right) = S_3$. In the paraxial approximation, $P_0$, 
$P_1$, $P_2$, and $P_3$ correspond to the Stokes parameters in the local coordinate frame $(\bm e_r, \bm e_{\varphi})$.
The presence of the terms in Eqs.~\eqref{jpolarpol} and~\eqref{jpolarph} which do not contain derivatives is due to the fact that 
the directions of the $\bm e_r$ and $\bm e_{\varphi}$ vectors depend on $\varphi$.

As a commonly used example, we consider the class of the Bessel beams which are characterized by the integer
index $m$ of the projection of the total angular momentum onto the beam axis~\cite{Knyazev2018, Maurer2007, Sederberg2020, Nesterov2000}. 
The electric field in the beam decomposed over the plane waves has the form 
\begin{equation}\label{decomposed} 
\boldsymbol{E}(\bm r,z) = E_0 \mathrm{e}^{\i q_z z} 
\sum\limits_{\bm q_{\parallel}} a(\bm q_{\parallel}) \exp \left(\mathrm{i}\bm q_{\parallel} \cdot \bm r \right)\bm e_{\bm q} \,,
\end{equation}
where $E_0$ is the amplitude, $\bm q = (\bm q_{\parallel}, q_z)$ is the wavevector with the in-plane component $\bm q_{\parallel}$,
$q = \omega n_{\omega} /c$, $n_{\omega}$ is the refractive index of the medium, $\bm e_{\bm q} \perp \bm q$ is the unit polarization vector
of the plane wave, and $a(\bm q_{\parallel})$ is the Fourier coefficient. The Bessel beams are formed from the plane waves with the 
wavevectors $\bm q$ lying on the surface of the cone with a certain angle $\theta_q$ and the axis parallel to $z$. Therefore, 
$q_z = q \cos \theta_q$ and $q_{\parallel} = q \sin \theta_q$ are fixed, and the integration in Eq.~\eqref{decomposed} is performed 
over the directions of the wavevector $\bm q_{\parallel}$. 

The vector $\bm e_{\bm q}$ determines the polarization (spin momentum) of the plane waves constituting the Bessel beam. 
We take $\bm e_{\bm q}$ in the form 
\begin{equation}
\label{polarization}
  \bm e_{\bm q} = \alpha \, \bm e_{\theta q} + \beta \, \bm e_{\varphi q} \:,
\end{equation} 
where $\alpha$ and $\beta$ are complex numbers, $|\alpha|^2 + |\beta|^2 = 1$,  
$\bm e_{\theta q} = (\cos \theta_q \cos \phi_q, \cos \theta_q \sin \phi_q, -\sin\theta_q)$ and $\bm e_{\phi q} = (-\sin \phi_q, \cos \phi_q, 0)$ 
are the unit vectors orthogonal to each other and to the wavevector $\bm q$. The cases of $\alpha = 1/\sqrt{2}$ and $\beta = \pm \i/\sqrt{2}$, for example, correspond to the beams composed of the circularly polarized plane waves.
The total angular momentum projection $m$ of the twisted field is determined 
by the dependence of $a(\bm q_{\parallel})$ on the polar angle $\varphi_q$ of the in-plane wavevector $\bm q_{\parallel}$, 
which is taken in the form 
\begin{equation}\label{a_q}
a(\bm q_{\parallel}) = \i^{-m-1}  (2\pi/ q_{\parallel})  \delta(q_{\parallel} - q \sin \theta_q) \exp(\i m \varphi_q) \,.
\end{equation}

By calculating the sum in Eq.~\eqref{decomposed} with $\bm e_{\bm q}$ and $a(\bm q_{\parallel})$ given by Eqs.~\eqref{polarization} and~\eqref{a_q}, respectively, one obtains the in-plane components of the electric field in the polar coordinate frame and in the paraxial approximation ($\theta_q \ll 1$)~\cite{Knyazev2018}
\begin{align}
\label{Etwisted1}
E_r(r,\varphi) = \frac{E_0}{2} \mathrm{e}^{\mathrm{i}m\varphi} \bigl[ o_+ J_{m+1}(q_{\parallel} r)  - o_- J_{m-1}(q_{\parallel} r)\bigr] \,, \nonumber \\
E_{\varphi}(r,\varphi) = \frac{E_0}{2\mathrm{i}} \mathrm{e}^{\mathrm{i}m\varphi} \bigl[o_+ J_{m+1}(q_{\parallel} r)
+ o_- J_{m-1}(q_{\parallel} r)\bigr] \,,
\end{align}
where $o_\pm = \alpha \pm \i \beta$ and $J_{m}$ is the Bessel function with the index $m$. In particular, $o_{\pm} = 1$
for the ``radial'' Bessel beam constructed from $p$-polarized plane waves ($\alpha = 1$, $\beta = 0$) and $o_{\pm} = \pm \i$
for the ``azimuthal'' Bessel beam constructed from $s$-polarized waves ($\alpha = 0$, $\beta = 1$). 
For the Bessel beams constructed 
from circularly polarized plane waves, only one of the coefficients, either $o_+$ or $o_-$, is nonzero.

The straightforward calculation of the photocurrent components~(\ref{jpolarth}--\ref{jpolarph}) with the electric field given by Eq.~\eqref{Etwisted1} yields $j_\varphi^{(\rm th)} = 0$,
\begin{eqnarray}\label{jth_bessel}
j_r^{(\rm th)} = j_0 \frac{\tau_\varepsilon}{\tau} \{ J_{m+1} (J_{m} - J_{m+2} ) - J_{m-1}(J_{m} - J_{m-2})  \nonumber \\
 - [J_{m+1} (J_{m} - J_{m+2} ) + J_{m-1}(J_{m} - J_{m-2})] \, p_3 \} \,, \;\;\; \;\;
\end{eqnarray}
\begin{eqnarray}
j_r^{(\rm pol)} +  j_r^{(\rm ph)} = j_0 J_m ( J_{m+1} - J_{m-1} ) \, p_1 \:, \;\;\;\;\;\;\;\;\;\;\;\; \nonumber 
\end{eqnarray}
\begin{eqnarray}\label{jpolph_bessel}
j_{\varphi}^{(\rm pol)} +  j_{\varphi}^{(\rm ph)} = j_0 J_m ( J_{m+1} - J_{m-1} ) \left( p_2 + \frac{p_3}{\omega \tau} \right)  \nonumber \\
- \frac{j_0}{\omega \tau} J_m ( J_{m+1} +  J_{m-1} ) , \;\;
\end{eqnarray}
where 
\begin{equation}
    j_0 = - \frac{n e^3\tau^3 E_0^2 q_{\parallel}}{ {m^*}^2 (1+\omega^2\tau^2)} \,,
\end{equation}
$p_1 = |\alpha|^2 - |\beta|^2$, $p_2 = \alpha \beta^* + \alpha^* \beta$, and $p_3 = \i(\alpha \beta^* - \alpha^* \beta)$ are the polarization parameters of the plane waves constituting the Bessel beam, and $J_m = J_m(q_{\parallel} r)$.

Now, we illustrate the photocurrents induced by Bessel beams with different polarizations and angular momentum projections. We focus on the polarization- and phase-sensitive photocurrents $\boldsymbol{j}^{(\mathrm{pol})} + \boldsymbol{j}^{(\mathrm{ph})}$. The photothermoelectric current $\boldsymbol{j}^{(\mathrm{th})}$ has only radial component and is determined solely by the gradient of the radiation intensity. Note that, for the beams composed of circularly or elliptically polarized plane waves ($p_3 \neq 0$), the spatial distribution of the radiation intensity and, hence, the photothermoelectric current  depend on the sign of circular polarization, see Eq.~\eqref{jth_bessel}.

\begin{figure}[hptb]
\includegraphics[scale=0.67]{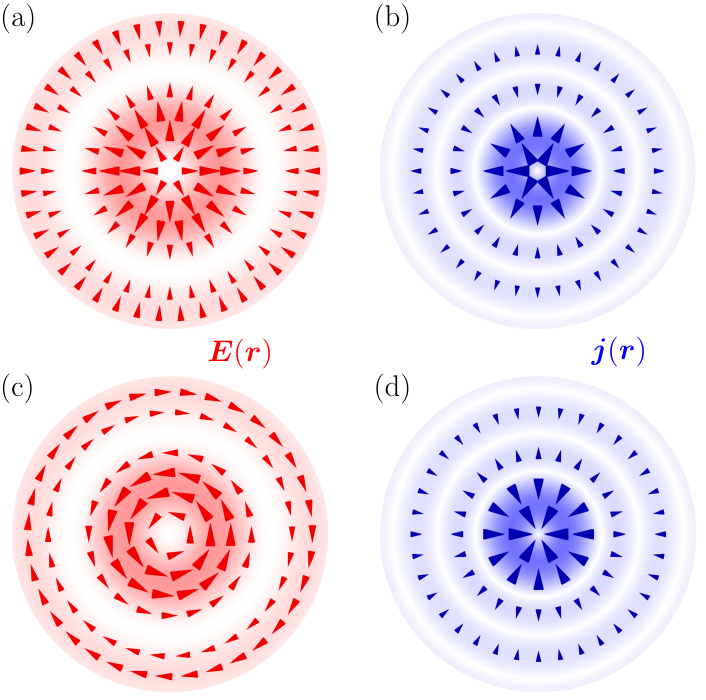}
\caption{Photocurrents induced by the radial and azimuthal Bessel beams with $m=0$. (a) and (b) Distributions of the electric field vector $\boldsymbol E(\boldsymbol r)$ and the photocurrent density $\boldsymbol{j}(\boldsymbol r) = \boldsymbol{j}^{(\mathrm{pol})} + \boldsymbol{j}^{(\mathrm{ph})}$ for the radial Bessel beam. (c) and (d) Distributions of the electric field vector $\boldsymbol E(\boldsymbol r)$ and the photocurrent $\boldsymbol{j}(\boldsymbol r)$ for the azimuthal Bessel beam. Red and blue backgrounds encode the radiation intensity and the magnitude of the photocurrent density, respectively.}
\label{Fig_m0_lin}
\end{figure}

Figure~\ref{Fig_m0_lin} shows the spatial distributions of the ac electric field and the photocurrent density for the radial ($p_1=1$) and azimuthal ($p_1=-1$) Bessel beams with the total angular momentum projection $m=0$. In these particular cases, the electric field in the beams has only radial or azimuthal component, and the field distribution is rotational invariant, Figs.~\ref{Fig_m0_lin}(a) and~\ref{Fig_m0_lin}(c). For the both beams, the photocurrent has only radial component, Figs.~\ref{Fig_m0_lin}(b) and~\ref{Fig_m0_lin}(d). The current density is non-monotonic and oscillates with the distance from the beam center, these oscillations originate from the radial distribution of the electric field. While the local current magnitude is the same for the radial and azimuthal Bessel beams, the current directions are opposite and controlled by the polarization of plane waves constituting the beams. Note, that the photothermoelectric effect also produces a radial electric current. This contribution, however, is independent of the linear polarization and, therefore, can be experimentally separated.

\begin{figure}[hptb]
\includegraphics[scale=0.67]{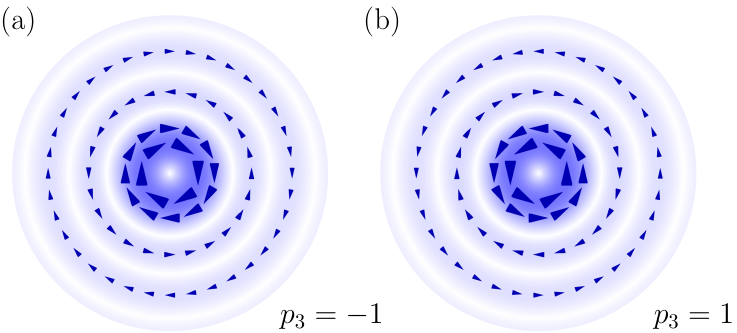}
\caption{Spatial distributions of the photocurrent density $\boldsymbol{j}(\boldsymbol r) = \boldsymbol{j}^{(\mathrm{pol})} + \boldsymbol{j}^{(\mathrm{ph})}$ for the Bessel beams with $m=0$ composed of (a) left-handed ($p_3=-1$) and (b) right-handed ($p_3=1$) circularly polarized plane waves. The current $\boldsymbol{j}$ forms vortices with the winding direction determined by the radiation helicity.}
\label{M0_circ}
\end{figure}

Whereas the photocurrents induced by the radial and azimuthal Bessel beams with $m=0$ flow radially, the photocurrents induced by the beams composed of circularly polarized plane waves ($p_3 = \pm 1$) have polarization-sensitive azimuthal components. This is shown in Fig.~\ref{M0_circ} for the Bessel beams with $m=0$. In these cases, the photocurrents form vortices (here, the sets of concentric current loops with alternating directions) with the winding direction determined by the radiation helicity.

\begin{figure}[hptb]
\includegraphics[scale=0.67]{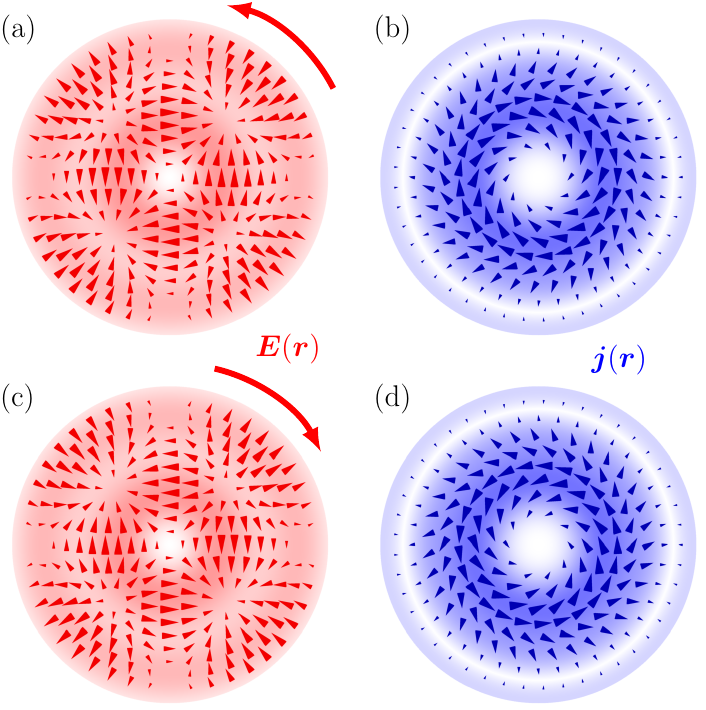}
\caption{Photocurrents induced by the radial Bessel beams with $m=\pm2$. (a) and (c) Snapshots of electric field distributions in the cross-sections of the beams. (b) and (d) Distributions of the photocurrent density $\boldsymbol{j}(\boldsymbol r) = \boldsymbol{j}^{(\mathrm{pol})} + \boldsymbol{j}^{(\mathrm{ph})}$ for the corresponding Bessel beams.}
\label{Mpm2_rad}
\end{figure}

For the Bessel beams with $m\neq0$, the spatial distributions of the ac field amplitude and the photocurrent density get more complex with the radial and azimuthal components intermixed.
Figures~\ref{Mpm2_rad}(a) and~\ref{Mpm2_rad}(c) show the snapshots of the electric field distributions in the radial ($p_1 = 1$) Bessel beams with $m=\pm2$ at a certain time $t_0$. The field distributions at any time $t$ are obtained by the rotation of the given distributions 
by the angle $\phi = \omega(t-t_0)/m$, as indicated by bent red arrows. Far from the beam center, the field is linearly polarized along the radius, whereas at the beam core the polarization is elliptical. Figures~\ref{Mpm2_rad}(b) and~\ref{Mpm2_rad}(d) show the distributions of the photocurrent density in the beams. The photocurrents contain both radial and azimuthal components. The direction of the azimuthal component is controlled by the sign of $m$ and, hence, it is opposite in Figs.~\ref{Mpm2_rad}(b) and~\ref{Mpm2_rad}(d).

\begin{figure}[hptb]
\includegraphics[width=\linewidth]{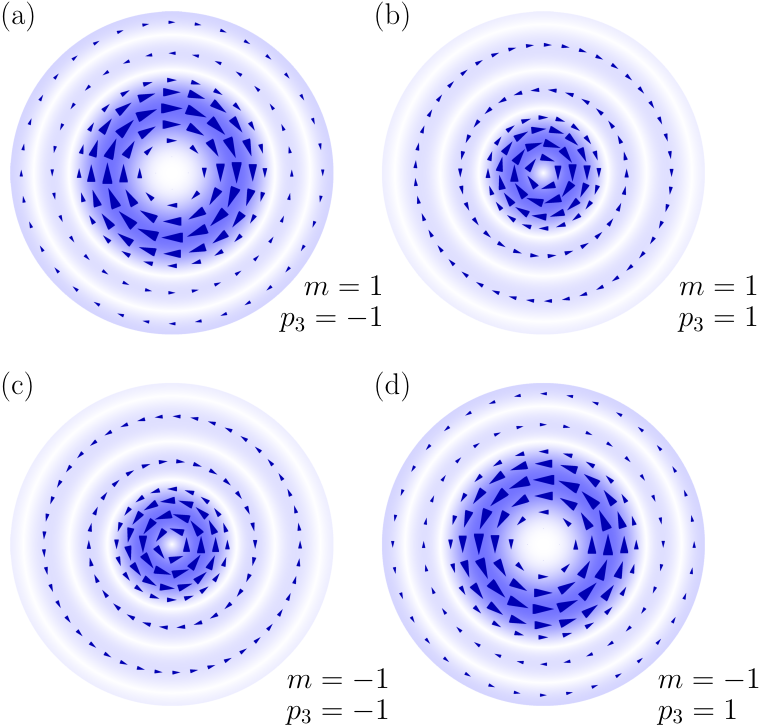}
\caption{Spatial distributions of the photocurrent density $\boldsymbol{j}(\boldsymbol r) = \boldsymbol{j}^{(\mathrm{pol})} + \boldsymbol{j}^{(\mathrm{ph})}$ for the Bessel beams with $m=\pm1$ composed of left-handed ($p_3=-1$) and right-handed ($p_3=1$) circularly polarized plane waves. (a) $m=1,\ p_3=-1$, (b) $m=1,\ p_3=1$, (c) $m=-1,\ p_3=-1$, and (d) $m=-1,\ p_3=1$.}
\label{l02_circ}
\end{figure}

Figure~\ref{l02_circ} shows the spatial distributions of the photocurrent density for the Bessel beams with $m=\pm1$ composed of the left-handed ($p_3=-1$) and the right-handed ($p_3=1$) circularly polarized plane waves. Similar to the case of $m=0$ (Fig.~\ref{M0_circ}), the photocurrents have polarization-sensitive azimuthal components and form vortices. The current distribution in the beam cross-section is determined by both the total angular momentum projection $m$ and the radiation helicity. Reversing the sign of both $m$ and $p_3$ mirrors the distributions, cf. Figs.~\ref{l02_circ}(a) and~\ref{l02_circ}(d) or Figs.~\ref{l02_circ}(b) and~\ref{l02_circ}(c).

\begin{figure}[hptb]
\includegraphics[scale=0.69]{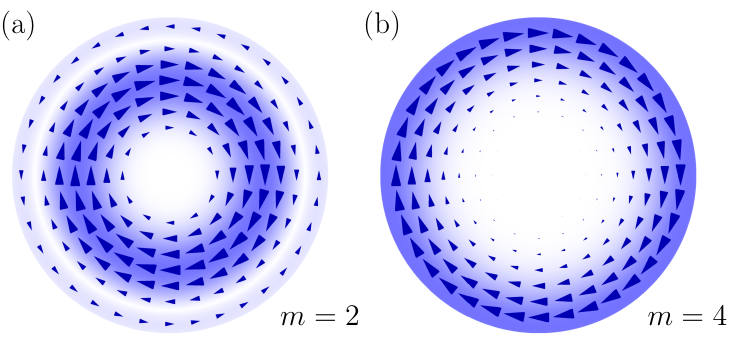}
\caption{Spatial distributions of the photocurrent density $\boldsymbol{j}(\boldsymbol r) = \boldsymbol{j}^{(\mathrm{pol})} + \boldsymbol{j}^{(\mathrm{ph})}$ for the Bessel beams with (a) $m=2$ and (b) $m=4$ composed of left-handed circularly polarized plane waves.}
\label{M24_circ}
\end{figure}

Completing the discussion of photocurrent distributions, we note that
the radius of the first circle in the Bessel beam increases almost linearly with the absolute
value of the orbital angular momentum projection $| l | = |m-p_3|$~\cite{Knyazev2018}. So does the radius of the first current loop. This is shown in Fig.~\ref{M24_circ}, where the spatial distributions of the current density are plotted for the Bessel beams with $m=2$ and $m=4$.

Radial photocurrents $j_r$ lead to a redistribution of electric charge in the 2DEG plane and, hence, to the (non-uniform) 
electrostatic potential $U(r)$, which, in turn, results in the radial drift current $j_{r}^{(\rm drift)} = - \sigma_0 \, d U / d r$, where $\sigma_0 = n e^2 \tau /m^*$ is the static conductivity. In the open circuit configuration, the total radial current in the steady-state regime vanishes, $j_r + j_{r}^{(\rm drift)} = 0$, 
which yields 
\begin{equation}\label{potential}
U(r) - U(0) = \frac{1}{\sigma_0}\int\limits_{0}^{r} j_r(r') dr' \,.
\end{equation}
Therefore, for the photocurrents induced by the Bessel beams, Eqs.~\eqref{jth_bessel} and~\eqref{jpolph_bessel}, we obtain 
\begin{eqnarray}\label{potential_final}
U(r) - U(0) =  U_0 \left[\delta_{m,0} - J^2_m(q_{\parallel} r)\right] p_1 \nonumber \\
+ U_0 \frac{\tau_\varepsilon}{\tau} \sum_{s = \pm 1} [J_{m+s}^2(q_{\parallel} r) - \delta_{m+s,0}](1 - s p_3) \,,
\end{eqnarray}
where
\begin{equation}
U_0 = - \frac{e \tau^2 E_0^2}{ m^* (1+\omega^2\tau^2)} \,,
\end{equation}
and $J_m$ is the Bessel function.
Interestingly, $U(\infty) - U(0)$ does not vanish for the beams with $m = 0, \pm 1$ only. This voltage is determined 
solely by the photocurrent $j_r^{(\rm pol)} +  j_r^{(\rm ph)}$ for the beams with $m=0$ and by the photothermoelectric current $j^{\rm (th)}_r$ 
for the beams with $m = \pm 1$. The photovoltage $U_0$ is 
of the order of $1$~mV for the terahertz beams with the intensity 1~kW/cm$^2$ and the structure parameters listed in the end of Sec.~III.

Azimuthal photocurrents $j_{\varphi}$ do not produce voltage drops in homogeneous 2D systems~\footnote{In samples with special geometry, like split rings, azimuthal photocurrents can also be detected as voltage drops}. What they produce is a static magnetic field $\bm B_{\rm photo}$, which points along the 2DEG plane normal at $z = 0$. At the center of the beam, the field can be readily calculated from the Biot-Savart law
\begin{equation}\label{Bfield} 
B_{\rm photo} = \frac{2\pi}{c}\int\limits_{0}^{+\infty}\frac{j_{\phi} (r)}{r} dr \,.
\end{equation}
For the azimuthal photocurrent given by Eq.~\eqref{jpolph_bessel}, we obtain 
\begin{equation}\label{B}
B_{\rm photo} = \frac{8 j_0}{c (4m^2-1)} \left( p_2 + \frac{p_3}{\omega \tau} + \frac{2 m}{\omega \tau} \right) \,,
\end{equation}
where $j_0 = - n e^3 \tau^3 E_0^2 q_{\parallel}/[m^{*2}(1+\omega^2 \tau^2)]$. The first two contributions in Eq.~\eqref{B} depend on the polarization parameters $p_2$ and $p_3$ of the plane waves constituting the beam.
The third contribution is polarization-independent but reverses its sign upon changing the sign of $m$. At large values 
of the total angular momentum projection, $|m|\gg 1$, the polarization-independent contribution dominates and the magnetic field 
assumes the dependence $B_{\rm photo} \propto 1/m$. Estimation of the current density for $q_\parallel/q = 0.1$ and the radiation intensity $1$~kW/cm$^2$ yields $j_0 \approx 30$~nA/cm. The corresponding magnetic field $B_{\rm photo}$ is of the order of nT. Such fields can be detected, e.g., by SQUIDs~\cite{Nowack2013} or optical sensors based on color centers in silicon carbide or diamond~\cite{Simin2016}.
The emergence of a static magnetization controlled by the angular momentum of the beam can be considered as the inverse Faraday effect~\cite{Durnev2023a} of twisted light.

\section{Conclusions}

In summary, we have studied the generation of direct electric currents in two-dimensional electron systems driven by electromagnetic radiation with structured intensity, polarization, and phase. The theory of such a non-linear and non-local response is developed within the kinetic approach for the intraband electron transport, which, for semiconductor structures, corresponds to the radiation of radio-frequency and terahertz spectral ranges. The derived analytical expressions for the photocurrent contributions are  general and can be applied to the radiation with an arbitrary spatial profile. In particular, the photocurrents are induced by the radiation with a uniform intensity but spatially varying polarization in the 2D plane, e.g., at the boundary between the domains excited by radiation with different polarization states. The total current emerging at the boundary of the domains excited by circularly polarized radiation with the opposite helicity flow along the boundary and does not depend on the boundary structure nor the electron gas mobility in the high-frequency limit. Its value for bilayer graphene is estimated as 20~$\mu$A for terahertz radiation with the intensity 1~kW/cm$^2$.

The photocurrents induced by the Bessel beams carrying orbital angular momentum have both the radial and azimuthal (vortex-like) components controlled by the beam polarization and angular momentum. In the experiments conducted in the open circuit configuration, the radial photocurrents lead to a charge redistribution in the sample and, hence, to a non-uniform electrostatic potential and oppositely directed drift currents. The amplitude of the resulting voltage sensitive to the beam polarization and angular momentum is of the order of 1~mV for the terahertz beams with the intensity 1~kW/cm$^2$. The azimuthal photocurrents, in turn, do not produce voltage drops in homogeneous 2D systems, however create a tiny static magnetic field directed perpendicular to the sample.
We have calculated the photoinduced electrostatic potential and the photoinduced magnetic field and analyzed their dependence on the beam parameters. The analysis suggests that the measurement of the photoresponse provides a useful experimental tool to determine the parameters of structured radiation, such as, e.g., the photon spin and orbital angular momentum.

\acknowledgments

This work was supported by the Russian Science Foundation (project No. 22-12-00211).

\bibliographystyle{apsrev4-1-customized}
\bibliography{bibliography}

\begin{thebibliography}{60}%
\makeatletter
\providecommand \@ifxundefined [1]{%
 \@ifx{#1\undefined}
}%
\providecommand \@ifnum [1]{%
 \ifnum #1\expandafter \@firstoftwo
 \else \expandafter \@secondoftwo
 \fi
}%
\providecommand \@ifx [1]{%
 \ifx #1\expandafter \@firstoftwo
 \else \expandafter \@secondoftwo
 \fi
}%
\providecommand \natexlab [1]{#1}%
\providecommand \enquote  [1]{``#1''}%
\providecommand \bibnamefont  [1]{#1}%
\providecommand \bibfnamefont [1]{#1}%
\providecommand \citenamefont [1]{#1}%
\providecommand \href@noop [0]{\@secondoftwo}%
\providecommand \href [0]{\begingroup \@sanitize@url \@href}%
\providecommand \@href[1]{\@@startlink{#1}\@@href}%
\providecommand \@@href[1]{\endgroup#1\@@endlink}%
\providecommand \@sanitize@url [0]{\catcode `\\12\catcode `\$12\catcode
  `\&12\catcode `\#12\catcode `\^12\catcode `\_12\catcode `\%12\relax}%
\providecommand \@@startlink[1]{}%
\providecommand \@@endlink[0]{}%
\providecommand \url  [0]{\begingroup\@sanitize@url \@url }%
\providecommand \@url [1]{\endgroup\@href {#1}{\urlprefix }}%
\providecommand \urlprefix  [0]{URL }%
\providecommand \Eprint [0]{\href }%
\providecommand \doibase [0]{http://dx.doi.org/}%
\providecommand \selectlanguage [0]{\@gobble}%
\providecommand \bibinfo  [0]{\@secondoftwo}%
\providecommand \bibfield  [0]{\@secondoftwo}%
\providecommand \translation [1]{[#1]}%
\providecommand \BibitemOpen [0]{}%
\providecommand \bibitemStop [0]{}%
\providecommand \bibitemNoStop [0]{.\EOS\space}%
\providecommand \EOS [0]{\spacefactor3000\relax}%
\providecommand \BibitemShut  [1]{\csname bibitem#1\endcsname}%
\let\auto@bib@innerbib\@empty
\bibitem [{\citenamefont {Forbes}\ \emph {et~al.}(2021)\citenamefont {Forbes},
  \citenamefont {de~Oliveira},\ and\ \citenamefont {Dennis}}]{Forbes2021}%
  \BibitemOpen
  \bibfield  {author} {\bibinfo {author} {\bibfnamefont {A.}~\bibnamefont
  {Forbes}}, \bibinfo {author} {\bibfnamefont {M.}~\bibnamefont {de~Oliveira}},
  \ and\ \bibinfo {author} {\bibfnamefont {M.~R.}\ \bibnamefont {Dennis}},\
  }{\bibinfo {title} {Structured light},}\ \href {\doibase
  10.1038/s41566-021-00780-4} {\bibfield  {journal} {\bibinfo  {journal} {Nat.
  Photonics}\ }\textbf {\bibinfo {volume} {15}},\ \bibinfo {pages} {253}
  (\bibinfo {year} {2021})}\BibitemShut {NoStop}%
\bibitem [{\citenamefont {Allen}\ \emph {et~al.}(1999)\citenamefont {Allen},
  \citenamefont {Padgett},\ and\ \citenamefont {Babiker}}]{Allen1999}%
  \BibitemOpen
  \bibfield  {author} {\bibinfo {author} {\bibfnamefont {L.}~\bibnamefont
  {Allen}}, \bibinfo {author} {\bibfnamefont {M.}~\bibnamefont {Padgett}}, \
  and\ \bibinfo {author} {\bibfnamefont {M.}~\bibnamefont {Babiker}},\ }\href
  {\doibase https://doi.org/10.1016/S0079-6638(08)70391-3} {\emph {\bibinfo
  {title} {IV The Orbital Angular Momentum of Light}}},\ Progress in Optics\
  (\bibinfo  {publisher} {Elsevier},\ \bibinfo {address} {Amsterdam},\ \bibinfo
  {year} {1999})\ pp.\ \bibinfo {pages} {291--372}\BibitemShut {NoStop}%
\bibitem [{\citenamefont {Molina-Terriza}\ \emph {et~al.}(2007)\citenamefont
  {Molina-Terriza}, \citenamefont {Torres},\ and\ \citenamefont
  {Torner}}]{MolinaTerriza2007}%
  \BibitemOpen
  \bibfield  {author} {\bibinfo {author} {\bibfnamefont {G.}~\bibnamefont
  {Molina-Terriza}}, \bibinfo {author} {\bibfnamefont {J.~P.}\ \bibnamefont
  {Torres}}, \ and\ \bibinfo {author} {\bibfnamefont {L.}~\bibnamefont
  {Torner}},\ }{\bibinfo {title} {Twisted photons},}\ \href {\doibase
  10.1038/nphys607} {\bibfield  {journal} {\bibinfo  {journal} {Nature Phys.}\
  }\textbf {\bibinfo {volume} {3}},\ \bibinfo {pages} {305} (\bibinfo {year}
  {2007})}\BibitemShut {NoStop}%
\bibitem [{\citenamefont {Knyazev}\ and\ \citenamefont
  {Serbo}(2018)}]{Knyazev2018}%
  \BibitemOpen
  \bibfield  {author} {\bibinfo {author} {\bibfnamefont {B.~A.}\ \bibnamefont
  {Knyazev}}\ and\ \bibinfo {author} {\bibfnamefont {V.~G.}\ \bibnamefont
  {Serbo}},\ }{\bibinfo {title} {Beams of photons with nonzero orbital angular
  momentum projection: new results},}\ \href
  {https://ufn.ru/en/articles/2018/5/e/} {\bibfield  {journal} {\bibinfo
  {journal} {Phys. Usp.}\ }\textbf {\bibinfo {volume} {61}},\ \bibinfo {pages}
  {449} (\bibinfo {year} {2018})}\BibitemShut {NoStop}%
\bibitem [{\citenamefont {Maurer}\ \emph {et~al.}(2007)\citenamefont {Maurer},
  \citenamefont {Jesacher}, \citenamefont {F{\"u}rhapter}, \citenamefont
  {Bernet},\ and\ \citenamefont {Ritsch-Marte}}]{Maurer2007}%
  \BibitemOpen
  \bibfield  {author} {\bibinfo {author} {\bibfnamefont {C.}~\bibnamefont
  {Maurer}}, \bibinfo {author} {\bibfnamefont {A.}~\bibnamefont {Jesacher}},
  \bibinfo {author} {\bibfnamefont {S.}~\bibnamefont {F{\"u}rhapter}}, \bibinfo
  {author} {\bibfnamefont {S.}~\bibnamefont {Bernet}}, \ and\ \bibinfo {author}
  {\bibfnamefont {M.}~\bibnamefont {Ritsch-Marte}},\ }{\bibinfo {title}
  {Tailoring of arbitrary optical vector beams},}\ \href {\doibase
  10.1088/1367-2630/9/3/078} {\bibfield  {journal} {\bibinfo  {journal} {New J.
  Phys.}\ }\textbf {\bibinfo {volume} {9}},\ \bibinfo {pages} {78} (\bibinfo
  {year} {2007})}\BibitemShut {NoStop}%
\bibitem [{\citenamefont {Wei}\ \emph {et~al.}(2015)\citenamefont {Wei},
  \citenamefont {Liu}, \citenamefont {Niu}, \citenamefont {Zhang},
  \citenamefont {Wang}, \citenamefont {Yang},\ and\ \citenamefont
  {Liu}}]{Wei2015}%
  \BibitemOpen
  \bibfield  {author} {\bibinfo {author} {\bibfnamefont {X.}~\bibnamefont
  {Wei}}, \bibinfo {author} {\bibfnamefont {C.}~\bibnamefont {Liu}}, \bibinfo
  {author} {\bibfnamefont {L.}~\bibnamefont {Niu}}, \bibinfo {author}
  {\bibfnamefont {Z.}~\bibnamefont {Zhang}}, \bibinfo {author} {\bibfnamefont
  {K.}~\bibnamefont {Wang}}, \bibinfo {author} {\bibfnamefont {Z.}~\bibnamefont
  {Yang}}, \ and\ \bibinfo {author} {\bibfnamefont {J.}~\bibnamefont {Liu}},\
  }{\bibinfo {title} {Generation of arbitrary order bessel beams via 3d printed
  axicons at the terahertz frequency range},}\ \href {\doibase
  10.1364/AO.54.010641} {\bibfield  {journal} {\bibinfo  {journal} {Appl.
  Opt.}\ }\textbf {\bibinfo {volume} {54}},\ \bibinfo {pages} {10641} (\bibinfo
  {year} {2015})}\BibitemShut {NoStop}%
\bibitem [{\citenamefont {Choporova}\ \emph {et~al.}(2017)\citenamefont
  {Choporova}, \citenamefont {Knyazev}, \citenamefont {Kulipanov},
  \citenamefont {Pavelyev}, \citenamefont {Scheglov}, \citenamefont
  {Vinokurov}, \citenamefont {Volodkin},\ and\ \citenamefont
  {Zhabin}}]{Choporova2017}%
  \BibitemOpen
  \bibfield  {author} {\bibinfo {author} {\bibfnamefont {Y.~Y.}\ \bibnamefont
  {Choporova}}, \bibinfo {author} {\bibfnamefont {B.~A.}\ \bibnamefont
  {Knyazev}}, \bibinfo {author} {\bibfnamefont {G.~N.}\ \bibnamefont
  {Kulipanov}}, \bibinfo {author} {\bibfnamefont {V.~S.}\ \bibnamefont
  {Pavelyev}}, \bibinfo {author} {\bibfnamefont {M.~A.}\ \bibnamefont
  {Scheglov}}, \bibinfo {author} {\bibfnamefont {N.~A.}\ \bibnamefont
  {Vinokurov}}, \bibinfo {author} {\bibfnamefont {B.~O.}\ \bibnamefont
  {Volodkin}}, \ and\ \bibinfo {author} {\bibfnamefont {V.~N.}\ \bibnamefont
  {Zhabin}},\ }{\bibinfo {title} {High-power bessel beams with orbital angular
  momentum in the terahertz range},}\ \href {\doibase
  10.1103/PhysRevA.96.023846} {\bibfield  {journal} {\bibinfo  {journal} {Phys.
  Rev. A}\ }\textbf {\bibinfo {volume} {96}},\ \bibinfo {pages} {023846}
  (\bibinfo {year} {2017})}\BibitemShut {NoStop}%
\bibitem [{\citenamefont {Zhifeng}\ \emph {et~al.}(2020)\citenamefont
  {Zhifeng}, \citenamefont {Xingdu}, \citenamefont {Bikashkali}, \citenamefont
  {Kevin}, \citenamefont {Jingbo}, \citenamefont {Tianwei}, \citenamefont
  {Wenjing}, \citenamefont {Ritesh}, \citenamefont {Miquel}, \citenamefont
  {Stefano}, \citenamefont {M.},\ and\ \citenamefont {Liang}}]{Zhifeng2020}%
  \BibitemOpen
  \bibfield  {author} {\bibinfo {author} {\bibfnamefont {Z.}~\bibnamefont
  {Zhifeng}}, \bibinfo {author} {\bibfnamefont {Q.}~\bibnamefont {Xingdu}},
  \bibinfo {author} {\bibfnamefont {M.}~\bibnamefont {Bikashkali}}, \bibinfo
  {author} {\bibfnamefont {L.}~\bibnamefont {Kevin}}, \bibinfo {author}
  {\bibfnamefont {S.}~\bibnamefont {Jingbo}}, \bibinfo {author} {\bibfnamefont
  {W.}~\bibnamefont {Tianwei}}, \bibinfo {author} {\bibfnamefont
  {L.}~\bibnamefont {Wenjing}}, \bibinfo {author} {\bibfnamefont
  {A.}~\bibnamefont {Ritesh}}, \bibinfo {author} {\bibfnamefont {J.~J.}\
  \bibnamefont {Miquel}}, \bibinfo {author} {\bibfnamefont {L.}~\bibnamefont
  {Stefano}}, \bibinfo {author} {\bibfnamefont {L.~N.}\ \bibnamefont {M.}}, \
  and\ \bibinfo {author} {\bibfnamefont {F.}~\bibnamefont {Liang}},\ }{\bibinfo
  {title} {Tunable topological charge vortex microlaser},}\ \href {\doibase
  10.1126/science.aba8996} {\bibfield  {journal} {\bibinfo  {journal}
  {Science}\ }\textbf {\bibinfo {volume} {368}},\ \bibinfo {pages} {760}
  (\bibinfo {year} {2020})}\BibitemShut {NoStop}%
\bibitem [{\citenamefont {Ashkin}(2000)}]{Ashkin2000}%
  \BibitemOpen
  \bibfield  {author} {\bibinfo {author} {\bibfnamefont {A.}~\bibnamefont
  {Ashkin}},\ }{\bibinfo {title} {History of optical trapping and manipulation
  of small-neutral particle, atoms, and molecules},}\ \href {\doibase
  10.1109/2944.902132} {\bibfield  {journal} {\bibinfo  {journal} {IEEE Journal
  of Selected Topics in Quantum Electronics}\ }\textbf {\bibinfo {volume}
  {6}},\ \bibinfo {pages} {841} (\bibinfo {year} {2000})}\BibitemShut {NoStop}%
\bibitem [{\citenamefont {Andrews}(2011)}]{Andrews_book2011}%
  \BibitemOpen
  \bibfield  {author} {\bibinfo {author} {\bibfnamefont {D.~L.}\ \bibnamefont
  {Andrews}},\ }\href@noop {} {\emph {\bibinfo {title} {Structured Light and
  Its Applications: An Introduction to Phase-Structured Beams and Nanoscale
  Optical Forces}}}\ (\bibinfo  {publisher} {Academic Press},\ \bibinfo
  {address} {Cambridge},\ \bibinfo {year} {2011})\BibitemShut {NoStop}%
\bibitem [{\citenamefont {Marag{\`o}}\ \emph {et~al.}(2013)\citenamefont
  {Marag{\`o}}, \citenamefont {Jones}, \citenamefont {Gucciardi}, \citenamefont
  {Volpe},\ and\ \citenamefont {Ferrari}}]{Marago2013}%
  \BibitemOpen
  \bibfield  {author} {\bibinfo {author} {\bibfnamefont {O.~M.}\ \bibnamefont
  {Marag{\`o}}}, \bibinfo {author} {\bibfnamefont {P.~H.}\ \bibnamefont
  {Jones}}, \bibinfo {author} {\bibfnamefont {P.~G.}\ \bibnamefont
  {Gucciardi}}, \bibinfo {author} {\bibfnamefont {G.}~\bibnamefont {Volpe}}, \
  and\ \bibinfo {author} {\bibfnamefont {A.~C.}\ \bibnamefont {Ferrari}},\
  }{\bibinfo {title} {Optical trapping and manipulation of nanostructures},}\
  \href {\doibase 10.1038/nnano.2013.208} {\bibfield  {journal} {\bibinfo
  {journal} {Nature Nanotech.}\ }\textbf {\bibinfo {volume} {8}},\ \bibinfo
  {pages} {807} (\bibinfo {year} {2013})}\BibitemShut {NoStop}%
\bibitem [{\citenamefont {Urban}\ \emph {et~al.}(2014)\citenamefont {Urban},
  \citenamefont {Carretero-Palacios}, \citenamefont {Lutich}, \citenamefont
  {Lohm{\"u}ller}, \citenamefont {Feldmann},\ and\ \citenamefont
  {J{\"a}ckel}}]{Urban2014}%
  \BibitemOpen
  \bibfield  {author} {\bibinfo {author} {\bibfnamefont {A.~S.}\ \bibnamefont
  {Urban}}, \bibinfo {author} {\bibfnamefont {S.}~\bibnamefont
  {Carretero-Palacios}}, \bibinfo {author} {\bibfnamefont {A.~A.}\ \bibnamefont
  {Lutich}}, \bibinfo {author} {\bibfnamefont {T.}~\bibnamefont
  {Lohm{\"u}ller}}, \bibinfo {author} {\bibfnamefont {J.}~\bibnamefont
  {Feldmann}}, \ and\ \bibinfo {author} {\bibfnamefont {F.}~\bibnamefont
  {J{\"a}ckel}},\ }{\bibinfo {title} {Optical trapping and manipulation of
  plasmonic nanoparticles: fundamentals{,} applications{,} and perspectives},}\
  \href {\doibase 10.1039/C3NR06617G} {\bibfield  {journal} {\bibinfo
  {journal} {Nanoscale}\ }\textbf {\bibinfo {volume} {6}},\ \bibinfo {pages}
  {4458} (\bibinfo {year} {2014})}\BibitemShut {NoStop}%
\bibitem [{\citenamefont {Mantsevich}\ and\ \citenamefont
  {Tarasenko}(2017)}]{Mantsevich2017}%
  \BibitemOpen
  \bibfield  {author} {\bibinfo {author} {\bibfnamefont {V.~N.}\ \bibnamefont
  {Mantsevich}}\ and\ \bibinfo {author} {\bibfnamefont {S.~A.}\ \bibnamefont
  {Tarasenko}},\ }{\bibinfo {title} {Fluid photonic crystal from colloidal
  quantum dots},}\ \href {\doibase 10.1103/PhysRevA.96.033855} {\bibfield
  {journal} {\bibinfo  {journal} {Phys. Rev. A}\ }\textbf {\bibinfo {volume}
  {96}},\ \bibinfo {pages} {033855} (\bibinfo {year} {2017})}\BibitemShut
  {NoStop}%
\bibitem [{\citenamefont {Yang}\ \emph {et~al.}(2021)\citenamefont {Yang},
  \citenamefont {Ren}, \citenamefont {Chen}, \citenamefont {Arita},\ and\
  \citenamefont {Rosales-Guzm{\'a}n}}]{Yang2021}%
  \BibitemOpen
  \bibfield  {author} {\bibinfo {author} {\bibfnamefont {Y.}~\bibnamefont
  {Yang}}, \bibinfo {author} {\bibfnamefont {Y.}~\bibnamefont {Ren}}, \bibinfo
  {author} {\bibfnamefont {M.}~\bibnamefont {Chen}}, \bibinfo {author}
  {\bibfnamefont {Y.}~\bibnamefont {Arita}}, \ and\ \bibinfo {author}
  {\bibfnamefont {C.}~\bibnamefont {Rosales-Guzm{\'a}n}},\ }{\bibinfo {title}
  {{Optical trapping with structured light: a review}},}\ \href {\doibase
  10.1117/1.AP.3.3.034001} {\bibfield  {journal} {\bibinfo  {journal} {Advanced
  Photonics}\ }\textbf {\bibinfo {volume} {3}},\ \bibinfo {pages} {034001}
  (\bibinfo {year} {2021})}\BibitemShut {NoStop}%
\bibitem [{\citenamefont {Ji}\ \emph {et~al.}(2020)\citenamefont {Ji},
  \citenamefont {Liu}, \citenamefont {Krylyuk}, \citenamefont {Fan},
  \citenamefont {Zhang}, \citenamefont {Pan}, \citenamefont {Feng},
  \citenamefont {Davydov},\ and\ \citenamefont {Agarwal}}]{Ji2020}%
  \BibitemOpen
  \bibfield  {author} {\bibinfo {author} {\bibfnamefont {Z.}~\bibnamefont
  {Ji}}, \bibinfo {author} {\bibfnamefont {W.}~\bibnamefont {Liu}}, \bibinfo
  {author} {\bibfnamefont {S.}~\bibnamefont {Krylyuk}}, \bibinfo {author}
  {\bibfnamefont {X.}~\bibnamefont {Fan}}, \bibinfo {author} {\bibfnamefont
  {Z.}~\bibnamefont {Zhang}}, \bibinfo {author} {\bibfnamefont
  {A.}~\bibnamefont {Pan}}, \bibinfo {author} {\bibfnamefont {L.}~\bibnamefont
  {Feng}}, \bibinfo {author} {\bibfnamefont {A.}~\bibnamefont {Davydov}}, \
  and\ \bibinfo {author} {\bibfnamefont {R.}~\bibnamefont {Agarwal}},\
  }{\bibinfo {title} {Photocurrent detection of the orbital angular momentum of
  light},}\ \href {\doibase 10.1126/science.aba9192} {\bibfield  {journal}
  {\bibinfo  {journal} {Science}\ }\textbf {\bibinfo {volume} {368}},\ \bibinfo
  {pages} {763} (\bibinfo {year} {2020})}\BibitemShut {NoStop}%
\bibitem [{\citenamefont {Sederberg}\ \emph {et~al.}(2020)\citenamefont
  {Sederberg}, \citenamefont {Kong}, \citenamefont {Hufnagel}, \citenamefont
  {Zhang}, \citenamefont {Karimi},\ and\ \citenamefont
  {Corkum}}]{Sederberg2020}%
  \BibitemOpen
  \bibfield  {author} {\bibinfo {author} {\bibfnamefont {S.}~\bibnamefont
  {Sederberg}}, \bibinfo {author} {\bibfnamefont {F.}~\bibnamefont {Kong}},
  \bibinfo {author} {\bibfnamefont {F.}~\bibnamefont {Hufnagel}}, \bibinfo
  {author} {\bibfnamefont {C.}~\bibnamefont {Zhang}}, \bibinfo {author}
  {\bibfnamefont {E.}~\bibnamefont {Karimi}}, \ and\ \bibinfo {author}
  {\bibfnamefont {P.~B.}\ \bibnamefont {Corkum}},\ }{\bibinfo {title}
  {Vectorized optoelectronic control and metrology in a semiconductor},}\ \href
  {\doibase 10.1038/s41566-020-0690-1} {\bibfield  {journal} {\bibinfo
  {journal} {Nat. Photonics}\ }\textbf {\bibinfo {volume} {14}},\ \bibinfo
  {pages} {680} (\bibinfo {year} {2020})}\BibitemShut {NoStop}%
\bibitem [{\citenamefont {Lai}\ \emph {et~al.}(2022)\citenamefont {Lai},
  \citenamefont {Ma}, \citenamefont {Fan}, \citenamefont {Song}, \citenamefont
  {Yu}, \citenamefont {Liu}, \citenamefont {Zhang}, \citenamefont {Shi},
  \citenamefont {Cheng},\ and\ \citenamefont {Sun}}]{Lai2022}%
  \BibitemOpen
  \bibfield  {author} {\bibinfo {author} {\bibfnamefont {J.}~\bibnamefont
  {Lai}}, \bibinfo {author} {\bibfnamefont {J.}~\bibnamefont {Ma}}, \bibinfo
  {author} {\bibfnamefont {Z.}~\bibnamefont {Fan}}, \bibinfo {author}
  {\bibfnamefont {X.}~\bibnamefont {Song}}, \bibinfo {author} {\bibfnamefont
  {P.}~\bibnamefont {Yu}}, \bibinfo {author} {\bibfnamefont {Z.}~\bibnamefont
  {Liu}}, \bibinfo {author} {\bibfnamefont {P.}~\bibnamefont {Zhang}}, \bibinfo
  {author} {\bibfnamefont {Y.}~\bibnamefont {Shi}}, \bibinfo {author}
  {\bibfnamefont {J.}~\bibnamefont {Cheng}}, \ and\ \bibinfo {author}
  {\bibfnamefont {D.}~\bibnamefont {Sun}},\ }{\bibinfo {title} {{Direct light
  orbital angular momentum detection in mid-infrared based on the
  \mbox{type-II} Weyl semimetal TaIrTe$_4$}},}\ \href {\doibase
  https://doi.org/10.1002/adma.202201229} {\bibfield  {journal} {\bibinfo
  {journal} {Adv. Mater.}\ }\textbf {\bibinfo {volume} {34}},\ \bibinfo {pages}
  {2201229} (\bibinfo {year} {2022})}\BibitemShut {NoStop}%
\bibitem [{\citenamefont {Bhattacharya}\ \emph {et~al.}(2022)\citenamefont
  {Bhattacharya}, \citenamefont {Chaudhary}, \citenamefont {Grass},
  \citenamefont {Johnson}, \citenamefont {Wall},\ and\ \citenamefont
  {Lewenstein}}]{Bhattacharya2022}%
  \BibitemOpen
  \bibfield  {author} {\bibinfo {author} {\bibfnamefont {U.}~\bibnamefont
  {Bhattacharya}}, \bibinfo {author} {\bibfnamefont {S.}~\bibnamefont
  {Chaudhary}}, \bibinfo {author} {\bibfnamefont {T.}~\bibnamefont {Grass}},
  \bibinfo {author} {\bibfnamefont {A.~S.}\ \bibnamefont {Johnson}}, \bibinfo
  {author} {\bibfnamefont {S.}~\bibnamefont {Wall}}, \ and\ \bibinfo {author}
  {\bibfnamefont {M.}~\bibnamefont {Lewenstein}},\ }{\bibinfo {title}
  {Fermionic chern insulator from twisted light with linear polarization},}\
  \href {\doibase 10.1103/PhysRevB.105.L081406} {\bibfield  {journal} {\bibinfo
   {journal} {Phys. Rev. B}\ }\textbf {\bibinfo {volume} {105}},\ \bibinfo
  {pages} {L081406} (\bibinfo {year} {2022})}\BibitemShut {NoStop}%
\bibitem [{\citenamefont {Meier}\ and\ \citenamefont
  {Zakharchenya}(1984)}]{OO_book}%
  \BibitemOpen
  \bibinfo {editor} {\bibfnamefont {F.}~\bibnamefont {Meier}}\ and\ \bibinfo
  {editor} {\bibfnamefont {B.~P.}\ \bibnamefont {Zakharchenya}},\ eds.,\
  \href@noop {} {\emph {\bibinfo {title} {Optical Orientation}}}\ (\bibinfo
  {publisher} {North Holland},\ \bibinfo {address} {Amsterdam},\ \bibinfo
  {year} {1984})\BibitemShut {NoStop}%
\bibitem [{\citenamefont {Hubmann}\ \emph {et~al.}(2020)\citenamefont
  {Hubmann}, \citenamefont {Budkin}, \citenamefont {Otteneder}, \citenamefont
  {But}, \citenamefont {Sacr\'e}, \citenamefont {Yahniuk}, \citenamefont
  {Diendorfer}, \citenamefont {Bel'kov}, \citenamefont {Kozlov}, \citenamefont
  {Mikhailov}, \citenamefont {Dvoretsky}, \citenamefont {Varavin},
  \citenamefont {Remesnik}, \citenamefont {Tarasenko}, \citenamefont {Knap},\
  and\ \citenamefont {Ganichev}}]{Hubmann2020}%
  \BibitemOpen
  \bibfield  {author} {\bibinfo {author} {\bibfnamefont {S.}~\bibnamefont
  {Hubmann}}, \bibinfo {author} {\bibfnamefont {G.~V.}\ \bibnamefont {Budkin}},
  \bibinfo {author} {\bibfnamefont {M.}~\bibnamefont {Otteneder}}, \bibinfo
  {author} {\bibfnamefont {D.}~\bibnamefont {But}}, \bibinfo {author}
  {\bibfnamefont {D.}~\bibnamefont {Sacr\'e}}, \bibinfo {author} {\bibfnamefont
  {I.}~\bibnamefont {Yahniuk}}, \bibinfo {author} {\bibfnamefont
  {K.}~\bibnamefont {Diendorfer}}, \bibinfo {author} {\bibfnamefont {V.~V.}\
  \bibnamefont {Bel'kov}}, \bibinfo {author} {\bibfnamefont {D.~A.}\
  \bibnamefont {Kozlov}}, \bibinfo {author} {\bibfnamefont {N.~N.}\
  \bibnamefont {Mikhailov}}, \bibinfo {author} {\bibfnamefont {S.~A.}\
  \bibnamefont {Dvoretsky}}, \bibinfo {author} {\bibfnamefont {V.~S.}\
  \bibnamefont {Varavin}}, \bibinfo {author} {\bibfnamefont {V.~G.}\
  \bibnamefont {Remesnik}}, \bibinfo {author} {\bibfnamefont {S.~A.}\
  \bibnamefont {Tarasenko}}, \bibinfo {author} {\bibfnamefont {W.}~\bibnamefont
  {Knap}}, \ and\ \bibinfo {author} {\bibfnamefont {S.~D.}\ \bibnamefont
  {Ganichev}},\ }{\bibinfo {title} {{Symmetry breaking and circular
  photogalvanic effect in epitaxial
  ${\mathrm{Cd}}_{x}{\mathrm{Hg}}_{1\ensuremath{-}x}\mathrm{Te}$ films}},}\
  \href {\doibase 10.1103/PhysRevMaterials.4.043607} {\bibfield  {journal}
  {\bibinfo  {journal} {Phys. Rev. Mater.}\ }\textbf {\bibinfo {volume} {4}},\
  \bibinfo {pages} {043607} (\bibinfo {year} {2020})}\BibitemShut {NoStop}%
\bibitem [{\citenamefont {Sturman}\ and\ \citenamefont
  {Fridkin}(1992)}]{Sturman_book}%
  \BibitemOpen
  \bibfield  {author} {\bibinfo {author} {\bibfnamefont {B.~I.}\ \bibnamefont
  {Sturman}}\ and\ \bibinfo {author} {\bibfnamefont {V.~M.}\ \bibnamefont
  {Fridkin}},\ }\href@noop {} {\emph {\bibinfo {title} {The Photovoltaic and
  Photorefractive Effects in Noncentrosymmetric Materials}}}\ (\bibinfo
  {publisher} {Gordon and Breach Science Publishers},\ \bibinfo {address}
  {Philadelphia},\ \bibinfo {year} {1992})\BibitemShut {NoStop}%
\bibitem [{\citenamefont {Ivchenko}(2005)}]{Ivchenko_book}%
  \BibitemOpen
  \bibfield  {author} {\bibinfo {author} {\bibfnamefont {E.~L.}\ \bibnamefont
  {Ivchenko}},\ }\href@noop {} {\emph {\bibinfo {title} {Optical Spectroscopy
  of Semiconductor Nanostructures}}}\ (\bibinfo  {publisher} {Alpha Science},\
  \bibinfo {address} {Oxford},\ \bibinfo {year} {2005})\BibitemShut {NoStop}%
\bibitem [{\citenamefont {Sipe}\ and\ \citenamefont
  {Shkrebtii}(2000)}]{Sipe2000}%
  \BibitemOpen
  \bibfield  {author} {\bibinfo {author} {\bibfnamefont {J.~E.}\ \bibnamefont
  {Sipe}}\ and\ \bibinfo {author} {\bibfnamefont {A.~I.}\ \bibnamefont
  {Shkrebtii}},\ }{\bibinfo {title} {Second-order optical response in
  semiconductors},}\ \href {\doibase 10.1103/PhysRevB.61.5337} {\bibfield
  {journal} {\bibinfo  {journal} {Phys. Rev. B}\ }\textbf {\bibinfo {volume}
  {61}},\ \bibinfo {pages} {5337} (\bibinfo {year} {2000})}\BibitemShut
  {NoStop}%
\bibitem [{\citenamefont {Tarasenko}(2011)}]{Tarasenko2011}%
  \BibitemOpen
  \bibfield  {author} {\bibinfo {author} {\bibfnamefont {S.~A.}\ \bibnamefont
  {Tarasenko}},\ }{\bibinfo {title} {Direct current driven by ac electric field
  in quantum wells},}\ \href {\doibase 10.1103/PhysRevB.83.035313} {\bibfield
  {journal} {\bibinfo  {journal} {Phys. Rev. B}\ }\textbf {\bibinfo {volume}
  {83}},\ \bibinfo {pages} {035313} (\bibinfo {year} {2011})}\BibitemShut
  {NoStop}%
\bibitem [{\citenamefont {Durnev}\ and\ \citenamefont
  {Tarasenko}(2019)}]{Durnev2019}%
  \BibitemOpen
  \bibfield  {author} {\bibinfo {author} {\bibfnamefont {M.~V.}\ \bibnamefont
  {Durnev}}\ and\ \bibinfo {author} {\bibfnamefont {S.~A.}\ \bibnamefont
  {Tarasenko}},\ }{\bibinfo {title} {{High-frequency nonlinear transport and
  photogalvanic effects in 2D topological insulators}},}\ \href {\doibase
  10.1002/andp.201800418} {\bibfield  {journal} {\bibinfo  {journal} {Ann.
  Phys.}\ }\textbf {\bibinfo {volume} {531}},\ \bibinfo {pages} {1800418}
  (\bibinfo {year} {2019})}\BibitemShut {NoStop}%
\bibitem [{\citenamefont {Candussio}\ \emph {et~al.}(2020)\citenamefont
  {Candussio}, \citenamefont {Durnev}, \citenamefont {Tarasenko}, \citenamefont
  {Yin}, \citenamefont {Keil}, \citenamefont {Yang}, \citenamefont {Son},
  \citenamefont {Mishchenko}, \citenamefont {Plank}, \citenamefont {Bel'kov},
  \citenamefont {Slizovskiy}, \citenamefont {Fal'ko},\ and\ \citenamefont
  {Ganichev}}]{Candussio2020}%
  \BibitemOpen
  \bibfield  {author} {\bibinfo {author} {\bibfnamefont {S.}~\bibnamefont
  {Candussio}}, \bibinfo {author} {\bibfnamefont {M.~V.}\ \bibnamefont
  {Durnev}}, \bibinfo {author} {\bibfnamefont {S.~A.}\ \bibnamefont
  {Tarasenko}}, \bibinfo {author} {\bibfnamefont {J.}~\bibnamefont {Yin}},
  \bibinfo {author} {\bibfnamefont {J.}~\bibnamefont {Keil}}, \bibinfo {author}
  {\bibfnamefont {Y.}~\bibnamefont {Yang}}, \bibinfo {author} {\bibfnamefont
  {S.-K.}\ \bibnamefont {Son}}, \bibinfo {author} {\bibfnamefont
  {A.}~\bibnamefont {Mishchenko}}, \bibinfo {author} {\bibfnamefont
  {H.}~\bibnamefont {Plank}}, \bibinfo {author} {\bibfnamefont {V.~V.}\
  \bibnamefont {Bel'kov}}, \bibinfo {author} {\bibfnamefont {S.}~\bibnamefont
  {Slizovskiy}}, \bibinfo {author} {\bibfnamefont {V.}~\bibnamefont {Fal'ko}},
  \ and\ \bibinfo {author} {\bibfnamefont {S.~D.}\ \bibnamefont {Ganichev}},\
  }{\bibinfo {title} {Edge photocurrent driven by terahertz electric field in
  bilayer graphene},}\ \href {\doibase 10.1103/PhysRevB.102.045406} {\bibfield
  {journal} {\bibinfo  {journal} {Phys. Rev. B}\ }\textbf {\bibinfo {volume}
  {102}},\ \bibinfo {pages} {045406} (\bibinfo {year} {2020})}\BibitemShut
  {NoStop}%
\bibitem [{\citenamefont {Durnev}\ and\ \citenamefont
  {Tarasenko}(2021)}]{Durnev2021b}%
  \BibitemOpen
  \bibfield  {author} {\bibinfo {author} {\bibfnamefont {M.~V.}\ \bibnamefont
  {Durnev}}\ and\ \bibinfo {author} {\bibfnamefont {S.~A.}\ \bibnamefont
  {Tarasenko}},\ }{\bibinfo {title} {Edge photogalvanic effect caused by
  optical alignment of carrier momenta in two-dimensional dirac materials},}\
  \href {\doibase 10.1103/PhysRevB.103.165411} {\bibfield  {journal} {\bibinfo
  {journal} {Phys. Rev. B}\ }\textbf {\bibinfo {volume} {103}},\ \bibinfo
  {pages} {165411} (\bibinfo {year} {2021})}\BibitemShut {NoStop}%
\bibitem [{\citenamefont {Steiner}\ \emph {et~al.}(2022)\citenamefont
  {Steiner}, \citenamefont {Andreev},\ and\ \citenamefont
  {Breitkreiz}}]{Steiner2022}%
  \BibitemOpen
  \bibfield  {author} {\bibinfo {author} {\bibfnamefont {J.~F.}\ \bibnamefont
  {Steiner}}, \bibinfo {author} {\bibfnamefont {A.~V.}\ \bibnamefont
  {Andreev}}, \ and\ \bibinfo {author} {\bibfnamefont {M.}~\bibnamefont
  {Breitkreiz}},\ }{\bibinfo {title} {{Surface photogalvanic effect in Weyl
  semimetals}},}\ \href {\doibase 10.1103/PhysRevResearch.4.023021} {\bibfield
  {journal} {\bibinfo  {journal} {Phys. Rev. Res.}\ }\textbf {\bibinfo {volume}
  {4}},\ \bibinfo {pages} {023021} (\bibinfo {year} {2022})}\BibitemShut
  {NoStop}%
\bibitem [{\citenamefont {Leppenen}\ and\ \citenamefont
  {Golub}(2023)}]{Leppenen2023}%
  \BibitemOpen
  \bibfield  {author} {\bibinfo {author} {\bibfnamefont {N.~V.}\ \bibnamefont
  {Leppenen}}\ and\ \bibinfo {author} {\bibfnamefont {L.~E.}\ \bibnamefont
  {Golub}},\ }{\bibinfo {title} {Linear photogalvanic effect in surface states
  of topological insulators},}\ \href {\doibase 10.1103/PhysRevB.107.L161403}
  {\bibfield  {journal} {\bibinfo  {journal} {Phys. Rev. B}\ }\textbf {\bibinfo
  {volume} {107}},\ \bibinfo {pages} {L161403} (\bibinfo {year}
  {2023})}\BibitemShut {NoStop}%
\bibitem [{\citenamefont {Parafilo}\ \emph {et~al.}(2022)\citenamefont
  {Parafilo}, \citenamefont {Boev}, \citenamefont {Kovalev},\ and\
  \citenamefont {Savenko}}]{Parafilo2022}%
  \BibitemOpen
  \bibfield  {author} {\bibinfo {author} {\bibfnamefont {A.~V.}\ \bibnamefont
  {Parafilo}}, \bibinfo {author} {\bibfnamefont {M.~V.}\ \bibnamefont {Boev}},
  \bibinfo {author} {\bibfnamefont {V.~M.}\ \bibnamefont {Kovalev}}, \ and\
  \bibinfo {author} {\bibfnamefont {I.~G.}\ \bibnamefont {Savenko}},\
  }{\bibinfo {title} {Photogalvanic transport in fluctuating {I}sing
  superconductors},}\ \href {\doibase 10.1103/PhysRevB.106.144502} {\bibfield
  {journal} {\bibinfo  {journal} {Phys. Rev. B}\ }\textbf {\bibinfo {volume}
  {106}},\ \bibinfo {pages} {144502} (\bibinfo {year} {2022})}\BibitemShut
  {NoStop}%
\bibitem [{\citenamefont {Quinteiro}\ and\ \citenamefont
  {Berakdar}(2009)}]{Quinteiro2009b}%
  \BibitemOpen
  \bibfield  {author} {\bibinfo {author} {\bibfnamefont {G.~F.}\ \bibnamefont
  {Quinteiro}}\ and\ \bibinfo {author} {\bibfnamefont {J.}~\bibnamefont
  {Berakdar}},\ }{\bibinfo {title} {Electric currents induced by twisted light
  in quantum rings},}\ \href {\doibase 10.1364/OE.17.020465} {\bibfield
  {journal} {\bibinfo  {journal} {Opt. Express}\ }\textbf {\bibinfo {volume}
  {17}},\ \bibinfo {pages} {20465} (\bibinfo {year} {2009})}\BibitemShut
  {NoStop}%
\bibitem [{\citenamefont {W{\"a}tzel}\ and\ \citenamefont
  {Berakdar}(2016)}]{Waetzel2016}%
  \BibitemOpen
  \bibfield  {author} {\bibinfo {author} {\bibfnamefont {J.}~\bibnamefont
  {W{\"a}tzel}}\ and\ \bibinfo {author} {\bibfnamefont {J.}~\bibnamefont
  {Berakdar}},\ }{\bibinfo {title} {Centrifugal photovoltaic and photogalvanic
  effects driven by structured light},}\ \href {\doibase 10.1038/srep21475}
  {\bibfield  {journal} {\bibinfo  {journal} {Sci. Rep.}\ }\textbf {\bibinfo
  {volume} {6}},\ \bibinfo {pages} {21475} (\bibinfo {year}
  {2016})}\BibitemShut {NoStop}%
\bibitem [{\citenamefont {W\"atzel}\ \emph {et~al.}(2020)\citenamefont
  {W\"atzel}, \citenamefont {Sherman},\ and\ \citenamefont
  {Berakdar}}]{Waetzel2020}%
  \BibitemOpen
  \bibfield  {author} {\bibinfo {author} {\bibfnamefont {J.}~\bibnamefont
  {W\"atzel}}, \bibinfo {author} {\bibfnamefont {E.~Y.}\ \bibnamefont
  {Sherman}}, \ and\ \bibinfo {author} {\bibfnamefont {J.}~\bibnamefont
  {Berakdar}},\ }{\bibinfo {title} {Nanostructures in structured light:
  Photoinduced spin and orbital electron dynamics},}\ \href {\doibase
  10.1103/PhysRevB.101.235304} {\bibfield  {journal} {\bibinfo  {journal}
  {Phys. Rev. B}\ }\textbf {\bibinfo {volume} {101}},\ \bibinfo {pages}
  {235304} (\bibinfo {year} {2020})}\BibitemShut {NoStop}%
\bibitem [{\citenamefont {Quinteiro}\ and\ \citenamefont
  {Tamborenea}(2010)}]{Quinteiro2010}%
  \BibitemOpen
  \bibfield  {author} {\bibinfo {author} {\bibfnamefont {G.~F.}\ \bibnamefont
  {Quinteiro}}\ and\ \bibinfo {author} {\bibfnamefont {P.~I.}\ \bibnamefont
  {Tamborenea}},\ }{\bibinfo {title} {Twisted-light-induced optical transitions
  in semiconductors: free-carrier quantum kinetics},}\ \href {\doibase
  10.1103/PhysRevB.82.125207} {\bibfield  {journal} {\bibinfo  {journal} {Phys.
  Rev. B}\ }\textbf {\bibinfo {volume} {82}},\ \bibinfo {pages} {125207}
  (\bibinfo {year} {2010})}\BibitemShut {NoStop}%
\bibitem [{\citenamefont {Danishevskii}\ \emph {et~al.}(1970)\citenamefont
  {Danishevskii}, \citenamefont {Kastalskii}, \citenamefont {Ryvkin},\ and\
  \citenamefont {Yaroshetskii}}]{Danishevskii1970}%
  \BibitemOpen
  \bibfield  {author} {\bibinfo {author} {\bibfnamefont {A.}~\bibnamefont
  {Danishevskii}}, \bibinfo {author} {\bibfnamefont {A.}~\bibnamefont
  {Kastalskii}}, \bibinfo {author} {\bibfnamefont {S.}~\bibnamefont {Ryvkin}},
  \ and\ \bibinfo {author} {\bibfnamefont {I.}~\bibnamefont {Yaroshetskii}},\
  }{\bibinfo {title} {Dragging of free carriers by photons in direct interband
  transitions in semiconductors},}\ \href@noop {} {\bibfield  {journal}
  {\bibinfo  {journal} {Sov. Phys. JETP}\ }\textbf {\bibinfo {volume} {31}},\
  \bibinfo {pages} {292} (\bibinfo {year} {1970})}\BibitemShut {NoStop}%
\bibitem [{\citenamefont {Gibson}\ \emph {et~al.}(1970)\citenamefont {Gibson},
  \citenamefont {Kimmitt},\ and\ \citenamefont {Walker}}]{Gibson1970}%
  \BibitemOpen
  \bibfield  {author} {\bibinfo {author} {\bibfnamefont {A.~F.}\ \bibnamefont
  {Gibson}}, \bibinfo {author} {\bibfnamefont {M.~F.}\ \bibnamefont {Kimmitt}},
  \ and\ \bibinfo {author} {\bibfnamefont {A.~C.}\ \bibnamefont {Walker}},\
  }{\bibinfo {title} {{Photon drag in germanium}},}\ \href {\doibase
  10.1063/1.1653315} {\bibfield  {journal} {\bibinfo  {journal} {Appl. Phys.
  Lett.}\ }\textbf {\bibinfo {volume} {17}},\ \bibinfo {pages} {75} (\bibinfo
  {year} {1970})}\BibitemShut {NoStop}%
\bibitem [{\citenamefont {Perel'}\ and\ \citenamefont
  {Pinskii}(1973)}]{Perel1973}%
  \BibitemOpen
  \bibfield  {author} {\bibinfo {author} {\bibfnamefont {V.~I.}\ \bibnamefont
  {Perel'}}\ and\ \bibinfo {author} {\bibfnamefont {Y.~M.}\ \bibnamefont
  {Pinskii}},\ }{\bibinfo {title} {Constant current in conducting media due to
  a high-frequency electron electromagnetic field},}\ \href@noop {} {\bibfield
  {journal} {\bibinfo  {journal} {Sov. Phys. Solid State}\ }\textbf {\bibinfo
  {volume} {15}},\ \bibinfo {pages} {688} (\bibinfo {year} {1973})}\BibitemShut
  {NoStop}%
\bibitem [{\citenamefont {Luryi}(1987)}]{Luryi1987}%
  \BibitemOpen
  \bibfield  {author} {\bibinfo {author} {\bibfnamefont {S.}~\bibnamefont
  {Luryi}},\ }{\bibinfo {title} {Photon-drag effect in intersubband absorption
  by a two-dimensional electron gas},}\ \href {\doibase
  10.1103/PhysRevLett.58.2263} {\bibfield  {journal} {\bibinfo  {journal}
  {Phys. Rev. Lett.}\ }\textbf {\bibinfo {volume} {58}},\ \bibinfo {pages}
  {2263} (\bibinfo {year} {1987})}\BibitemShut {NoStop}%
\bibitem [{\citenamefont {Shalygin}\ \emph {et~al.}(2007)\citenamefont
  {Shalygin}, \citenamefont {Diehl}, \citenamefont {Hoffmann}, \citenamefont
  {Danilov}, \citenamefont {Herrle}, \citenamefont {Tarasenko}, \citenamefont
  {Schuh}, \citenamefont {Gerl}, \citenamefont {Wegscheider}, \citenamefont
  {Prettl},\ and\ \citenamefont {Ganichev}}]{Shalygin2007}%
  \BibitemOpen
  \bibfield  {author} {\bibinfo {author} {\bibfnamefont {V.~A.}\ \bibnamefont
  {Shalygin}}, \bibinfo {author} {\bibfnamefont {H.}~\bibnamefont {Diehl}},
  \bibinfo {author} {\bibfnamefont {C.}~\bibnamefont {Hoffmann}}, \bibinfo
  {author} {\bibfnamefont {S.~N.}\ \bibnamefont {Danilov}}, \bibinfo {author}
  {\bibfnamefont {T.}~\bibnamefont {Herrle}}, \bibinfo {author} {\bibfnamefont
  {S.~A.}\ \bibnamefont {Tarasenko}}, \bibinfo {author} {\bibfnamefont
  {D.}~\bibnamefont {Schuh}}, \bibinfo {author} {\bibfnamefont
  {C.}~\bibnamefont {Gerl}}, \bibinfo {author} {\bibfnamefont {W.}~\bibnamefont
  {Wegscheider}}, \bibinfo {author} {\bibfnamefont {W.}~\bibnamefont {Prettl}},
  \ and\ \bibinfo {author} {\bibfnamefont {S.~D.}\ \bibnamefont {Ganichev}},\
  }{\bibinfo {title} {Spin photocurrents and the circular photon drag effect in
  (110)-grown quantum well structures},}\ \href {\doibase
  10.1134/S0021364006220097} {\bibfield  {journal} {\bibinfo  {journal} {JETP
  Letters}\ }\textbf {\bibinfo {volume} {84}},\ \bibinfo {pages} {570}
  (\bibinfo {year} {2007})}\BibitemShut {NoStop}%
\bibitem [{\citenamefont {Hatano}\ \emph {et~al.}(2009)\citenamefont {Hatano},
  \citenamefont {Ishihara}, \citenamefont {Tikhodeev},\ and\ \citenamefont
  {Gippius}}]{Hatano2009}%
  \BibitemOpen
  \bibfield  {author} {\bibinfo {author} {\bibfnamefont {T.}~\bibnamefont
  {Hatano}}, \bibinfo {author} {\bibfnamefont {T.}~\bibnamefont {Ishihara}},
  \bibinfo {author} {\bibfnamefont {S.~G.}\ \bibnamefont {Tikhodeev}}, \ and\
  \bibinfo {author} {\bibfnamefont {N.~A.}\ \bibnamefont {Gippius}},\
  }{\bibinfo {title} {Transverse photovoltage induced by circularly polarized
  light},}\ \href {\doibase 10.1103/PhysRevLett.103.103906} {\bibfield
  {journal} {\bibinfo  {journal} {Phys. Rev. Lett.}\ }\textbf {\bibinfo
  {volume} {103}},\ \bibinfo {pages} {103906} (\bibinfo {year}
  {2009})}\BibitemShut {NoStop}%
\bibitem [{\citenamefont {Karch}\ \emph {et~al.}(2010)\citenamefont {Karch},
  \citenamefont {Olbrich}, \citenamefont {Schmalzbauer}, \citenamefont {Zoth},
  \citenamefont {Brinsteiner}, \citenamefont {Fehrenbacher}, \citenamefont
  {Wurstbauer}, \citenamefont {Glazov}, \citenamefont {Tarasenko},
  \citenamefont {Ivchenko}, \citenamefont {Weiss}, \citenamefont {Eroms},
  \citenamefont {Yakimova}, \citenamefont {Lara-Avila}, \citenamefont
  {Kubatkin},\ and\ \citenamefont {Ganichev}}]{Karch2010}%
  \BibitemOpen
  \bibfield  {author} {\bibinfo {author} {\bibfnamefont {J.}~\bibnamefont
  {Karch}}, \bibinfo {author} {\bibfnamefont {P.}~\bibnamefont {Olbrich}},
  \bibinfo {author} {\bibfnamefont {M.}~\bibnamefont {Schmalzbauer}}, \bibinfo
  {author} {\bibfnamefont {C.}~\bibnamefont {Zoth}}, \bibinfo {author}
  {\bibfnamefont {C.}~\bibnamefont {Brinsteiner}}, \bibinfo {author}
  {\bibfnamefont {M.}~\bibnamefont {Fehrenbacher}}, \bibinfo {author}
  {\bibfnamefont {U.}~\bibnamefont {Wurstbauer}}, \bibinfo {author}
  {\bibfnamefont {M.~M.}\ \bibnamefont {Glazov}}, \bibinfo {author}
  {\bibfnamefont {S.~A.}\ \bibnamefont {Tarasenko}}, \bibinfo {author}
  {\bibfnamefont {E.~L.}\ \bibnamefont {Ivchenko}}, \bibinfo {author}
  {\bibfnamefont {D.}~\bibnamefont {Weiss}}, \bibinfo {author} {\bibfnamefont
  {J.}~\bibnamefont {Eroms}}, \bibinfo {author} {\bibfnamefont
  {R.}~\bibnamefont {Yakimova}}, \bibinfo {author} {\bibfnamefont
  {S.}~\bibnamefont {Lara-Avila}}, \bibinfo {author} {\bibfnamefont
  {S.}~\bibnamefont {Kubatkin}}, \ and\ \bibinfo {author} {\bibfnamefont
  {S.~D.}\ \bibnamefont {Ganichev}},\ }{\bibinfo {title} {Dynamic hall effect
  driven by circularly polarized light in a graphene layer},}\ \href {\doibase
  10.1103/PhysRevLett.105.227402} {\bibfield  {journal} {\bibinfo  {journal}
  {Phys. Rev. Lett.}\ }\textbf {\bibinfo {volume} {105}},\ \bibinfo {pages}
  {227402} (\bibinfo {year} {2010})}\BibitemShut {NoStop}%
\bibitem [{\citenamefont {Entin}\ \emph {et~al.}(2010)\citenamefont {Entin},
  \citenamefont {Magarill},\ and\ \citenamefont {Shepelyansky}}]{Entin2010}%
  \BibitemOpen
  \bibfield  {author} {\bibinfo {author} {\bibfnamefont {M.~V.}\ \bibnamefont
  {Entin}}, \bibinfo {author} {\bibfnamefont {L.~I.}\ \bibnamefont {Magarill}},
  \ and\ \bibinfo {author} {\bibfnamefont {D.~L.}\ \bibnamefont
  {Shepelyansky}},\ }{\bibinfo {title} {Theory of resonant photon drag in
  monolayer graphene},}\ \href {\doibase 10.1103/PhysRevB.81.165441} {\bibfield
   {journal} {\bibinfo  {journal} {Phys. Rev. B}\ }\textbf {\bibinfo {volume}
  {81}},\ \bibinfo {pages} {165441} (\bibinfo {year} {2010})}\BibitemShut
  {NoStop}%
\bibitem [{\citenamefont {Stachel}\ \emph {et~al.}(2014)\citenamefont
  {Stachel}, \citenamefont {Budkin}, \citenamefont {Hagner}, \citenamefont
  {Bel'kov}, \citenamefont {Glazov}, \citenamefont {Tarasenko}, \citenamefont
  {Clowes}, \citenamefont {Ashley}, \citenamefont {Gilbertson},\ and\
  \citenamefont {Ganichev}}]{Stachel2014}%
  \BibitemOpen
  \bibfield  {author} {\bibinfo {author} {\bibfnamefont {S.}~\bibnamefont
  {Stachel}}, \bibinfo {author} {\bibfnamefont {G.~V.}\ \bibnamefont {Budkin}},
  \bibinfo {author} {\bibfnamefont {U.}~\bibnamefont {Hagner}}, \bibinfo
  {author} {\bibfnamefont {V.~V.}\ \bibnamefont {Bel'kov}}, \bibinfo {author}
  {\bibfnamefont {M.~M.}\ \bibnamefont {Glazov}}, \bibinfo {author}
  {\bibfnamefont {S.~A.}\ \bibnamefont {Tarasenko}}, \bibinfo {author}
  {\bibfnamefont {S.~K.}\ \bibnamefont {Clowes}}, \bibinfo {author}
  {\bibfnamefont {T.}~\bibnamefont {Ashley}}, \bibinfo {author} {\bibfnamefont
  {A.~M.}\ \bibnamefont {Gilbertson}}, \ and\ \bibinfo {author} {\bibfnamefont
  {S.~D.}\ \bibnamefont {Ganichev}},\ }{\bibinfo {title}
  {{Cyclotron-resonance-assisted photon drag effect in InSb/InAlSb quantum
  wells excited by terahertz radiation}},}\ \href {\doibase
  10.1103/PhysRevB.89.115435} {\bibfield  {journal} {\bibinfo  {journal} {Phys.
  Rev. B}\ }\textbf {\bibinfo {volume} {89}},\ \bibinfo {pages} {115435}
  (\bibinfo {year} {2014})}\BibitemShut {NoStop}%
\bibitem [{\citenamefont {Obraztsov}\ \emph {et~al.}(2014)\citenamefont
  {Obraztsov}, \citenamefont {Kanda}, \citenamefont {Konishi}, \citenamefont
  {Kuwata-Gonokami}, \citenamefont {Garnov}, \citenamefont {Obraztsov},\ and\
  \citenamefont {Svirko}}]{Obraztsov2014}%
  \BibitemOpen
  \bibfield  {author} {\bibinfo {author} {\bibfnamefont {P.~A.}\ \bibnamefont
  {Obraztsov}}, \bibinfo {author} {\bibfnamefont {N.}~\bibnamefont {Kanda}},
  \bibinfo {author} {\bibfnamefont {K.}~\bibnamefont {Konishi}}, \bibinfo
  {author} {\bibfnamefont {M.}~\bibnamefont {Kuwata-Gonokami}}, \bibinfo
  {author} {\bibfnamefont {S.~V.}\ \bibnamefont {Garnov}}, \bibinfo {author}
  {\bibfnamefont {A.~N.}\ \bibnamefont {Obraztsov}}, \ and\ \bibinfo {author}
  {\bibfnamefont {Y.~P.}\ \bibnamefont {Svirko}},\ }{\bibinfo {title}
  {Photon-drag-induced terahertz emission from graphene},}\ \href {\doibase
  10.1103/PhysRevB.90.241416} {\bibfield  {journal} {\bibinfo  {journal} {Phys.
  Rev. B}\ }\textbf {\bibinfo {volume} {90}},\ \bibinfo {pages} {241416}
  (\bibinfo {year} {2014})}\BibitemShut {NoStop}%
\bibitem [{\citenamefont {Glazov}\ and\ \citenamefont
  {Ganichev}(2014)}]{Glazov2014}%
  \BibitemOpen
  \bibfield  {author} {\bibinfo {author} {\bibfnamefont {M.}~\bibnamefont
  {Glazov}}\ and\ \bibinfo {author} {\bibfnamefont {S.}~\bibnamefont
  {Ganichev}},\ }{\bibinfo {title} {High frequency electric field induced
  nonlinear effects in graphene},}\ \href {\doibase
  https://doi.org/10.1016/j.physrep.2013.10.003} {\bibfield  {journal}
  {\bibinfo  {journal} {Phys. Rep.}\ }\textbf {\bibinfo {volume} {535}},\
  \bibinfo {pages} {101} (\bibinfo {year} {2014})}\BibitemShut {NoStop}%
\bibitem [{\citenamefont {Plank}\ \emph {et~al.}(2016)\citenamefont {Plank},
  \citenamefont {Golub}, \citenamefont {Bauer}, \citenamefont {Bel'kov},
  \citenamefont {Herrmann}, \citenamefont {Olbrich}, \citenamefont {Eschbach},
  \citenamefont {Plucinski}, \citenamefont {Schneider}, \citenamefont
  {Kampmeier}, \citenamefont {Lanius}, \citenamefont {Mussler}, \citenamefont
  {Gr\"utzmacher},\ and\ \citenamefont {Ganichev}}]{Plank2016}%
  \BibitemOpen
  \bibfield  {author} {\bibinfo {author} {\bibfnamefont {H.}~\bibnamefont
  {Plank}}, \bibinfo {author} {\bibfnamefont {L.~E.}\ \bibnamefont {Golub}},
  \bibinfo {author} {\bibfnamefont {S.}~\bibnamefont {Bauer}}, \bibinfo
  {author} {\bibfnamefont {V.~V.}\ \bibnamefont {Bel'kov}}, \bibinfo {author}
  {\bibfnamefont {T.}~\bibnamefont {Herrmann}}, \bibinfo {author}
  {\bibfnamefont {P.}~\bibnamefont {Olbrich}}, \bibinfo {author} {\bibfnamefont
  {M.}~\bibnamefont {Eschbach}}, \bibinfo {author} {\bibfnamefont
  {L.}~\bibnamefont {Plucinski}}, \bibinfo {author} {\bibfnamefont {C.~M.}\
  \bibnamefont {Schneider}}, \bibinfo {author} {\bibfnamefont {J.}~\bibnamefont
  {Kampmeier}}, \bibinfo {author} {\bibfnamefont {M.}~\bibnamefont {Lanius}},
  \bibinfo {author} {\bibfnamefont {G.}~\bibnamefont {Mussler}}, \bibinfo
  {author} {\bibfnamefont {D.}~\bibnamefont {Gr\"utzmacher}}, \ and\ \bibinfo
  {author} {\bibfnamefont {S.~D.}\ \bibnamefont {Ganichev}},\ }{\bibinfo
  {title} {{Photon drag effect in
  $({\mathrm{Bi}}_{1\ensuremath{-}x}{\mathrm{Sb}}_{x}){}_{2}{\mathrm{Te}}_{3}$
  three-dimensional topological insulators}},}\ \href {\doibase
  10.1103/PhysRevB.93.125434} {\bibfield  {journal} {\bibinfo  {journal} {Phys.
  Rev. B}\ }\textbf {\bibinfo {volume} {93}},\ \bibinfo {pages} {125434}
  (\bibinfo {year} {2016})}\BibitemShut {NoStop}%
\bibitem [{\citenamefont {Mikheev}\ \emph {et~al.}(2018)\citenamefont
  {Mikheev}, \citenamefont {Saushin}, \citenamefont {Styapshin},\ and\
  \citenamefont {Svirko}}]{Mikheev2018}%
  \BibitemOpen
  \bibfield  {author} {\bibinfo {author} {\bibfnamefont {G.~M.}\ \bibnamefont
  {Mikheev}}, \bibinfo {author} {\bibfnamefont {A.~S.}\ \bibnamefont
  {Saushin}}, \bibinfo {author} {\bibfnamefont {V.~M.}\ \bibnamefont
  {Styapshin}}, \ and\ \bibinfo {author} {\bibfnamefont {Y.~P.}\ \bibnamefont
  {Svirko}},\ }{\bibinfo {title} {Interplay of the photon drag and the surface
  photogalvanic effects in the metal-semiconductor nanocomposite},}\ \href
  {\doibase 10.1038/s41598-018-26923-2} {\bibfield  {journal} {\bibinfo
  {journal} {Sci. Rep.}\ }\textbf {\bibinfo {volume} {8}},\ \bibinfo {pages}
  {8644} (\bibinfo {year} {2018})}\BibitemShut {NoStop}%
\bibitem [{\citenamefont {Shi}\ \emph {et~al.}(2021)\citenamefont {Shi},
  \citenamefont {Zhang}, \citenamefont {Chang},\ and\ \citenamefont
  {Song}}]{Shi2021}%
  \BibitemOpen
  \bibfield  {author} {\bibinfo {author} {\bibfnamefont {L.-k.}\ \bibnamefont
  {Shi}}, \bibinfo {author} {\bibfnamefont {D.}~\bibnamefont {Zhang}}, \bibinfo
  {author} {\bibfnamefont {K.}~\bibnamefont {Chang}}, \ and\ \bibinfo {author}
  {\bibfnamefont {J.~C.~W.}\ \bibnamefont {Song}},\ }{\bibinfo {title}
  {Geometric photon-drag effect and nonlinear shift current in centrosymmetric
  crystals},}\ \href {\doibase 10.1103/PhysRevLett.126.197402} {\bibfield
  {journal} {\bibinfo  {journal} {Phys. Rev. Lett.}\ }\textbf {\bibinfo
  {volume} {126}},\ \bibinfo {pages} {197402} (\bibinfo {year}
  {2021})}\BibitemShut {NoStop}%
\bibitem [{\citenamefont {Kitagawa}\ \emph {et~al.}(2011)\citenamefont
  {Kitagawa}, \citenamefont {Oka}, \citenamefont {Brataas}, \citenamefont
  {Fu},\ and\ \citenamefont {Demler}}]{Kitagawa2011}%
  \BibitemOpen
  \bibfield  {author} {\bibinfo {author} {\bibfnamefont {T.}~\bibnamefont
  {Kitagawa}}, \bibinfo {author} {\bibfnamefont {T.}~\bibnamefont {Oka}},
  \bibinfo {author} {\bibfnamefont {A.}~\bibnamefont {Brataas}}, \bibinfo
  {author} {\bibfnamefont {L.}~\bibnamefont {Fu}}, \ and\ \bibinfo {author}
  {\bibfnamefont {E.}~\bibnamefont {Demler}},\ }{\bibinfo {title} {Transport
  properties of nonequilibrium systems under the application of light:
  photoinduced quantum {H}all insulators without {L}andau levels},}\ \href
  {\doibase 10.1103/PhysRevB.84.235108} {\bibfield  {journal} {\bibinfo
  {journal} {Phys. Rev. B}\ }\textbf {\bibinfo {volume} {84}},\ \bibinfo
  {pages} {235108} (\bibinfo {year} {2011})}\BibitemShut {NoStop}%
\bibitem [{\citenamefont {Lindner}\ \emph {et~al.}(2011)\citenamefont
  {Lindner}, \citenamefont {Refael},\ and\ \citenamefont
  {Galitski}}]{Lindner2011}%
  \BibitemOpen
  \bibfield  {author} {\bibinfo {author} {\bibfnamefont {N.~H.}\ \bibnamefont
  {Lindner}}, \bibinfo {author} {\bibfnamefont {G.}~\bibnamefont {Refael}}, \
  and\ \bibinfo {author} {\bibfnamefont {V.}~\bibnamefont {Galitski}},\
  }{\bibinfo {title} {Floquet topological insulator in semiconductor quantum
  wells},}\ \href {\doibase 10.1038/nphys1926} {\bibfield  {journal} {\bibinfo
  {journal} {Nature Phys.}\ }\textbf {\bibinfo {volume} {7}},\ \bibinfo {pages}
  {490} (\bibinfo {year} {2011})}\BibitemShut {NoStop}%
\bibitem [{\citenamefont {Lindhard}(1954)}]{Lindhard1954}%
  \BibitemOpen
  \bibfield  {author} {\bibinfo {author} {\bibfnamefont {J.}~\bibnamefont
  {Lindhard}},\ }{\bibinfo {title} {On the properties of a gas of charged
  particles},}\ \href@noop {} {\bibfield  {journal} {\bibinfo  {journal} {Kgl.
  Danske Videnskab. Selskab Mat.-Fys. Medd.}\ }\textbf {\bibinfo {volume} {28}}
  (\bibinfo {year} {1954})}\BibitemShut {NoStop}%
\bibitem [{\citenamefont {Giuliani}\ and\ \citenamefont
  {Vignale}(2005)}]{GiulianiVignale_book2005}%
  \BibitemOpen
  \bibfield  {author} {\bibinfo {author} {\bibfnamefont {G.}~\bibnamefont
  {Giuliani}}\ and\ \bibinfo {author} {\bibfnamefont {G.}~\bibnamefont
  {Vignale}},\ }\href@noop {} {\emph {\bibinfo {title} {Quantum Theory of the
  Electron Liquid}}}\ (\bibinfo  {publisher} {Cambridge University Press},\
  \bibinfo {address} {Cambridge},\ \bibinfo {year} {2005})\BibitemShut
  {NoStop}%
\bibitem [{Note1()}]{Note1}%
  \BibitemOpen
  \bibinfo {note} {See paragraph after Eq.~\protect \eqref {currentnew} for
  details}\BibitemShut {NoStop}%
\bibitem [{Note2()}]{Note2}%
  \BibitemOpen
  \bibinfo {note} {To describe the energy relaxation of electrons one needs to
  go beyond the collision integral~\protect \eqref
  {collision_integral}}\BibitemShut {NoStop}%
\bibitem [{\citenamefont {Durnev}\ and\ \citenamefont
  {Tarasenko}(2023)}]{Durnev2023}%
  \BibitemOpen
  \bibfield  {author} {\bibinfo {author} {\bibfnamefont {M.~V.}\ \bibnamefont
  {Durnev}}\ and\ \bibinfo {author} {\bibfnamefont {S.~A.}\ \bibnamefont
  {Tarasenko}},\ }{\bibinfo {title} {Edge currents induced by ac electric field
  in two-dimensional {D}irac structures},}\ \href {\doibase
  10.3390/app13074080} {\bibfield  {journal} {\bibinfo  {journal} {Appl. Sci.}\
  }\textbf {\bibinfo {volume} {13}},\ \bibinfo {pages} {4080} (\bibinfo {year}
  {2023})}\BibitemShut {NoStop}%
\bibitem [{\citenamefont {Nesterov}\ and\ \citenamefont
  {Niziev}(2000)}]{Nesterov2000}%
  \BibitemOpen
  \bibfield  {author} {\bibinfo {author} {\bibfnamefont {A.~V.}\ \bibnamefont
  {Nesterov}}\ and\ \bibinfo {author} {\bibfnamefont {V.~G.}\ \bibnamefont
  {Niziev}},\ }{\bibinfo {title} {Laser beams with axially symmetric
  polarization},}\ \href {\doibase 10.1088/0022-3727/33/15/310} {\bibfield
  {journal} {\bibinfo  {journal} {J. Phys. D: Appl. Phys.}\ }\textbf {\bibinfo
  {volume} {33}},\ \bibinfo {pages} {1817} (\bibinfo {year}
  {2000})}\BibitemShut {NoStop}%
\bibitem [{Note3()}]{Note3}%
  \BibitemOpen
  \bibinfo {note} {In samples with special geometry, like split rings,
  azimuthal photocurrents can also be detected as voltage drops}\BibitemShut
  {NoStop}%
\bibitem [{\citenamefont {Nowack}\ \emph {et~al.}(2013)\citenamefont {Nowack},
  \citenamefont {Spanton}, \citenamefont {Baenninger}, \citenamefont
  {K{\"o}nig}, \citenamefont {Kirtley}, \citenamefont {Kalisky}, \citenamefont
  {Ames}, \citenamefont {Leubner}, \citenamefont {Br{\"u}ne}, \citenamefont
  {Buhmann}, \citenamefont {Molenkamp}, \citenamefont {Goldhaber-Gordon},\ and\
  \citenamefont {Moler}}]{Nowack2013}%
  \BibitemOpen
  \bibfield  {author} {\bibinfo {author} {\bibfnamefont {K.~C.}\ \bibnamefont
  {Nowack}}, \bibinfo {author} {\bibfnamefont {E.~M.}\ \bibnamefont {Spanton}},
  \bibinfo {author} {\bibfnamefont {M.}~\bibnamefont {Baenninger}}, \bibinfo
  {author} {\bibfnamefont {M.}~\bibnamefont {K{\"o}nig}}, \bibinfo {author}
  {\bibfnamefont {J.~R.}\ \bibnamefont {Kirtley}}, \bibinfo {author}
  {\bibfnamefont {B.}~\bibnamefont {Kalisky}}, \bibinfo {author} {\bibfnamefont
  {C.}~\bibnamefont {Ames}}, \bibinfo {author} {\bibfnamefont {P.}~\bibnamefont
  {Leubner}}, \bibinfo {author} {\bibfnamefont {C.}~\bibnamefont {Br{\"u}ne}},
  \bibinfo {author} {\bibfnamefont {H.}~\bibnamefont {Buhmann}}, \bibinfo
  {author} {\bibfnamefont {L.~W.}\ \bibnamefont {Molenkamp}}, \bibinfo {author}
  {\bibfnamefont {D.}~\bibnamefont {Goldhaber-Gordon}}, \ and\ \bibinfo
  {author} {\bibfnamefont {K.~A.}\ \bibnamefont {Moler}},\ }{\bibinfo {title}
  {Imaging currents in hgte quantum wells in the quantum spin hall regime},}\
  \href {http://dx.doi.org/10.1038/nmat3682} {\bibfield  {journal} {\bibinfo
  {journal} {Nat Mater}\ }\textbf {\bibinfo {volume} {12}},\ \bibinfo {pages}
  {787} (\bibinfo {year} {2013})}\BibitemShut {NoStop}%
\bibitem [{\citenamefont {Simin}\ \emph {et~al.}(2016)\citenamefont {Simin},
  \citenamefont {Soltamov}, \citenamefont {Poshakinskiy}, \citenamefont
  {Anisimov}, \citenamefont {Babunts}, \citenamefont {Tolmachev}, \citenamefont
  {Mokhov}, \citenamefont {Trupke}, \citenamefont {Tarasenko}, \citenamefont
  {Sperlich}, \citenamefont {Baranov}, \citenamefont {Dyakonov},\ and\
  \citenamefont {Astakhov}}]{Simin2016}%
  \BibitemOpen
  \bibfield  {author} {\bibinfo {author} {\bibfnamefont {D.}~\bibnamefont
  {Simin}}, \bibinfo {author} {\bibfnamefont {V.~A.}\ \bibnamefont {Soltamov}},
  \bibinfo {author} {\bibfnamefont {A.~V.}\ \bibnamefont {Poshakinskiy}},
  \bibinfo {author} {\bibfnamefont {A.~N.}\ \bibnamefont {Anisimov}}, \bibinfo
  {author} {\bibfnamefont {R.~A.}\ \bibnamefont {Babunts}}, \bibinfo {author}
  {\bibfnamefont {D.~O.}\ \bibnamefont {Tolmachev}}, \bibinfo {author}
  {\bibfnamefont {E.~N.}\ \bibnamefont {Mokhov}}, \bibinfo {author}
  {\bibfnamefont {M.}~\bibnamefont {Trupke}}, \bibinfo {author} {\bibfnamefont
  {S.~A.}\ \bibnamefont {Tarasenko}}, \bibinfo {author} {\bibfnamefont
  {A.}~\bibnamefont {Sperlich}}, \bibinfo {author} {\bibfnamefont {P.~G.}\
  \bibnamefont {Baranov}}, \bibinfo {author} {\bibfnamefont {V.}~\bibnamefont
  {Dyakonov}}, \ and\ \bibinfo {author} {\bibfnamefont {G.~V.}\ \bibnamefont
  {Astakhov}},\ }{\bibinfo {title} {All-optical dc nanotesla magnetometry using
  silicon vacancy fine structure in isotopically purified silicon carbide},}\
  \href {\doibase 10.1103/PhysRevX.6.031014} {\bibfield  {journal} {\bibinfo
  {journal} {Phys. Rev. X}\ }\textbf {\bibinfo {volume} {6}},\ \bibinfo {pages}
  {031014} (\bibinfo {year} {2016})}\BibitemShut {NoStop}%
\bibitem [{\citenamefont {Durnev}(2023)}]{Durnev2023a}%
  \BibitemOpen
  \bibfield  {author} {\bibinfo {author} {\bibfnamefont {M.~V.}\ \bibnamefont
  {Durnev}},\ }{\bibinfo {title} {Faraday and {K}err rotation due to
  photoinduced orbital magnetization in a two-dimensional electron gas},}\
  \href {\doibase 10.1103/PhysRevB.108.125418} {\bibfield  {journal} {\bibinfo
  {journal} {Phys. Rev. B}\ }\textbf {\bibinfo {volume} {108}},\ \bibinfo
  {pages} {125418} (\bibinfo {year} {2023})}\BibitemShut {NoStop}%
\end{thebibliography}%

\end{document}